\documentclass[11pt,a4paper]{article}
\usepackage[margin=3.5cm]{geometry}

\usepackage{amsmath,amssymb}
\numberwithin{equation}{section}
\usepackage{stackrel}
\usepackage{graphicx}
\usepackage{slashed}
\usepackage{epstopdf}
\usepackage{color}

\begin{document}

\title{Spectral Functions in QFT\footnote{Notes of a series of talks given in February and April, 2015 at CEFyMAP, UNACH (Tuxtla Guti\'errez, M\'exico) and at IFLP, UNLP/CONICET (La Plata, Argentina) as part of the project ME/13/16 (MINCyT and CONACYT).}}
\author{Pablo Pisani\footnote{pisani@fisica.unlp.edu.ar}\\[2mm]
{\small Instituto de F\'\i sica La Plata, CONICET-UNLP}\\
{\small C.C.\ 67 (1900) La Plata, Argentina}}
\date{}

\maketitle

\begin{abstract}
We present a pedagogical exposition of some applications of functional methods in quantum field theory: we use heat-kernel and zeta-function techniques to study the Casimir effect, the pair production in strong electric fields, quantum fields at finite temperature and beta-functions for a self-interacting scalar field, QED and pure Yang-Mills theories. The more recent application to the UV/IR mixing phenomenon in noncommutative theories is also discussed in this framework.
\end{abstract}

\section{Introduction}

These lecture notes are intended to illustrate in a simple manner how spectral functions, viz.\ heat-trace and $\zeta$-function, are used to compute leading quantum corrections to physical quantities in field theory. The applications we consider are well-known and have been studied with one approach or another in most textbooks in Quantum Field Theory. Our main purpose is to readily provide, through a few plain examples, the basic tools of functional methods in field theory.

Quantum effects in a particular field theory can be read from the spectrum of a certain differential operator. The effective action, for example, is given by the determinant on the infinite-dimensional Hilbert space of an unbounded operator. This ``infinite'' determinant can be written in terms of a {\it divergent series}. Fortunately, several criteria exist which allow us to unambiguously determine the sum of some divergent series. This procedure of assigning a finite value to a divergent series is equivalent to the ``regularization'' of the UV-divergencies that occur in the computation of one-loop Feynman diagrams. The functional method described in the present notes defines the value of a divergent series in terms of analytic extensions in the complex plane, the so-called ``$\zeta$-function regularization''. The heat-kernel (or the corresponding heat-trace) is a mathematical tool closely related to the $\zeta$-function which serves as a complementary regularization scheme.

After introducing the heat-trace, the $\zeta$-function and the main relation between them (section \ref{defs}), we give an example where a $\zeta$-function is used to determine the value of some particular divergent series (section \ref{ae}). To show a physical consequence of this definition, we study the Casimir effect in its simplest version (sections \ref{cas1} and \ref{cas2}). We also give an interpretation of this definition of divergent sums by comparing the $\zeta$-function method with the regularization in terms of a general cutoff function (section \ref{rama}). Some relation between the use of analytic extensions and classical definitions of divergent series (Ces\`aro, Abel and Borel summability) is illustrated with examples in appendix \ref{divsums}.

Next we introduce one of the most powerful results in the theory of spectral functions: the asymptotic expansion of the heat-trace (section \ref{ht}); this expansion allows us to write a general expression for the effective action in terms of the Seeley-de Witt coefficients. As an example, we compute the $\beta$-function of a real scalar field with a quartic self-interaction (section \ref{efac}). The equivalence between the heat-trace and the $\zeta$-function regularizations is then understood from the point of view of renormalization theory in QFT (section \ref{rela}).

The heat-trace can be used, in particular, to study the effect of quantum oscillations of an electron/positron system in constant magnetic or electric backgrounds. In this context, we compute the $\beta$-function of QED and the rate of pair production in the presence of strong electric fields (section \ref{sch}).

Our last examples consist in the computation of the $\beta$-function of Yang-Mills theories through the straightforward application of the heat-trace asymptotic expansion (section \ref{ym}), of thermodynamical quantities in field theory at finite temperature (section \ref{fintem}) and the propagator in a noncommutative field theory, whose nonplanar contribution exhibits the UV/IR mixing effect (section \ref{nc}).

For the sake of clarity, we have chosen the simplest cases leaving aside any complexity which could obscure the presentation. Furthermore, the main mathematical results are stated without a detailed description of the assumed hypotheses, or a rigourous mathematical proof. As a compensation for these omissions we expect to provide --through quite explicit examples-- the tools for a succinct calculation of these well-known quantum effects in field theory.

\section{Heat-trace and $\zeta$-function}\label{defs}

Heat-kernel and $\zeta$-function techniques deal with the calculation of the following related sums:
\begin{align}
  \sum_n e^{-\tau\,\lambda_n}
  \qquad{\rm and}\qquad \sum_n \lambda_n^{-s}\,,
\end{align}
where $\tau>0$, $s\in\mathbb{C}$ with $\mathcal{R}(s)$ large enough and $\lambda_n$ are eigenvalues of some positive definite (differential) operator, i.e.,
\begin{align}
  A\,\phi_n=\lambda_n\,\phi_n
  \qquad {\rm with\ } \lambda_n> 0\,.
\end{align}
In general, $n$ represents a multi-index or even a continuous variable that characterizes the eigenstate. Of particular interest are the expansion of the first sum (heat-trace) for small $\tau$ and the analytic extension of the second sum ($\zeta$-function) to the whole complex $s$-plane.

Let us begin by considering a second order differential operator $A$ acting on a Hilbert space of functions $\mathcal{H}$; take, for example, the one-dimensional Laplacian
\begin{align}
  A:=-\partial^2_x
\end{align}
acting on the space $\mathcal{H}= L_2(0,1)$ of square integrable functions $\phi(x)$ on the interval $[0,1]$ with Dirichlet boundary conditions $\phi(0)=\phi(1)=0$.

Let us then assume that $\mathcal{H}$ has an orthonormal basis of eigenfunctions $\phi_1(x),$\ $\phi_2(x),\ldots$ of $A$ with eigenvalues $\lambda_1,\lambda_2,\ldots$ In our example, the normalized eigenfunctions of the one-dimensional Laplacian
\begin{align}
  \phi_n(x)=\sqrt{2}\, \sin{(n\pi x)}
  \qquad {\rm with\ }n=1,2,3,\ldots\,,
\end{align}
have eigenvalues
\begin{align}
  \lambda_n=\pi^2n^2\,,
\end{align}
which, as expected, grow to infinity with $n$.

The heat-trace can now be defined as
\begin{align}
  {\rm Tr}\,e^{-\tau A}=\sum_n e^{-\tau\,\lambda_n}\,,
\end{align}
for $\tau>0$. In our example
\begin{align}
  {\rm Tr}\,e^{-\tau (-\partial^2_x)}=\sum_{n=1}^\infty e^{-\tau\,\pi^2n^2}\,,
\end{align}
which can be written in terms of a Jacobi theta-function,
\begin{align}
  \sum_{n=1}^\infty e^{-\tau\,\pi^2n^2}&=\tfrac12\,\vartheta_3(0|i\pi\tau)-\tfrac12\,.
\end{align}
Since the operator of our example does not have a zero mode, the heat-trace decreases exponentially as $\tau\rightarrow +\infty$. More important to us is that, due to the infinite dimensionality of the Hilbert space, the heat-trace diverges for small $\tau$; the general behaviour of the heat-trace for small $\tau$ is established in section \eqref{ht}.

The $\zeta$-function is defined as
\begin{align}
  \zeta(s)={\rm Tr}\,A^{-s}=\sum_n \lambda_n^{-s}
\end{align}
if $\mathcal{R}(s)$ is large enough, or as its (unique) analytic extension if not. In our example,
\begin{align}
  \zeta(s)=\pi^{-2s}\,\zeta_R(2s)\,,
\end{align}
where $\zeta_R$ is known as the Riemann $\zeta$-function, and defined as
\begin{align}\label{rie}
  \zeta_R(s)=\sum_{n=1}^\infty n^{-s}\,,
\end{align}
for $\mathcal{R}(s)>1$, or as its analytic extension otherwise. As a function on the whole complex plane $\zeta_R(s)$ has a unique simple pole at $s=1$ and vanishes at negative even integers; B.\ Riemann conjectured\footnote{``Hiervon w\"are allerdings ein strenger Beweis zu w\"unschen; ich habe indess die Aufsuchung desselben nach einigen fl\"uchtigen vergeblichen Versuchen vorl\"aufig bei Seite gelassen, da er f\"ur den n\"achsten Zweck meiner Untersuchung entbehrlich schien'' \cite{Riemann}.} that any other zero of $\zeta_R(s)$ must be located at the line $\mathcal{R}(s)=\tfrac12$.

We can now derive a very useful relation between both spectral functions. Applying the relation (Mellin transform)
\begin{align}
  \lambda^{-s}=\frac{1}{\Gamma(s)}\int_0^\infty d\tau\,\tau^{s-1}\,e^{-\tau\,\lambda}
\end{align}
to each eigenvalue $\lambda_n$ and performing the sum over the whole spectrum of $A$ we obtain
\begin{align}\label{meltra}
  \zeta(s)=\frac{1}{\Gamma(s)}\int_0^\infty d\tau\,\tau^{s-1}\,{\rm Tr}\,e^{-\tau A}\,.
\end{align}
This expression allows us to compute the analytic extension of the $\zeta$-function from the asymptotic expansion of the heat-trace for small values of $\tau$.

\section{Example: the Riemann $\zeta$-function}\label{ae}

As an example of the concepts of the previous section, let us consider the case in which the eigenvalues of the operator $A$ are given by $\lambda_n=n=1,2,3,\ldots$ The corresponding $\zeta$-function is then the Riemann $\zeta$-function $\zeta_R(s)$, defined as the analytic extension of the series given by \eqref{rie}, which is convergent for $\mathcal{R}(s)>1$. In order to study its analytic extension to the whole complex plane we compute the corresponding heat-trace
\begin{align}
  {\rm Tr}\,e^{-\tau A}=\sum_{n=1}^\infty e^{-\tau\,n}=\frac{1}{e^{\tau}-1}
\end{align}
and use Mellin transform, given by \eqref{meltra}, to write
\begin{align}
  \zeta_R(s)=\frac{1}{\Gamma(s)}\int_0^\infty d\tau\,\tau^{s-1}\,\frac{1}{e^{\tau}-1}\,.
\end{align}
As expected, the integral is well-defined only for $\mathcal{R}(s)>1$; otherwise it diverges at $\tau=0$. In fact, the analytic extension of $\zeta_R(s)$ to the rest of the complex plane is determined by the behaviour of the heat-trace for small values of $\tau$. Therefore, we separate the integral at large values of $\tau$ (say $\tau>1$) from small values of $\tau$ (then $0<\tau<1$) and use for the latter the expansion \cite{GR}
\begin{align}
  \frac{\tau}{e^\tau-1}&=\sum_{n=0}^\infty B_n\,\frac{\tau^n}{n!}
  =1-\frac12\,\tau+\frac1{6}\,\frac{\tau^2}{2!}-\frac{1}{30}\,\frac{\tau^4}{4!}
  +\frac{1}{42}\,\frac{\tau^6}{6!}-\ldots
\end{align}
which is actually valid for $|\tau|<2\pi$; $B_n$ are the Bernoulli numbers. After integrating this powers series in the interval $[0,1]$ we obtain
\begin{align}\label{finrie}
  \zeta_R(s)
  &=\frac{1}{\Gamma(s)}\left\{\int_1^\infty d\tau\,\frac{\tau^{s-1}}{e^{\tau}-1}
  +\sum_{n=0}^\infty \frac{B_n}{n!}\,\frac{1}{s+n-1}\right\}\,.
\end{align}
Although the previous manipulations are only valid for $\mathcal{R}(s)>1$, expression \eqref{finrie} now gives the analytic extension of $\zeta_R(s)$ to the whole complex plane: the first term between brackets is an entire function whereas the series in the second term is analytic for all $s\in\mathbb{C}$, except where it presents simple poles --namely, at $s=1$ and at all negative integers. However, since the term $1/\Gamma(s)$ vanishes at negative integers then the Riemann $\zeta$-function has a unique pole at $s=1$, the residue being $B_0=1$.

The term $1/\Gamma(s)$ also facilitates the determination of $\zeta_R(s)$ at each negative integer, which is given by the corresponding residue of the expression in brackets. For instance, at $s=0$ only the term in the series corresponding to $n=1$ gives a nonvanishing value,
\begin{align}\label{zeta0}
  \zeta_R(0)&=B_1=-\frac12\,.
\end{align}
Since the value of $\zeta_R(s)$ at a negative integer $-k$ is proportional to $B_{k+1}$ and Bernoulli numbers of odd order vanish --except from $B_1$-- then the Riemann $\zeta$-function vanishes at even negative integers; these are the trivial zeroes of $\zeta_R(s)$. On the contrary, at odd negative integers $\zeta_R(-2k+1)=-B_{2k}/2k$, so that, e.g.,
\begin{align}\label{zeta-1}
  \zeta_R(-1)&=-\frac{B_2}{2}=-\frac1{12}\,.
\end{align}

In summary, expression \eqref{rie} is only valid for $\mathcal{R}(s)>1$ but it suffices to define $\zeta_R(s)$ in the whole complex plane as its unique analytic extension. In the $\zeta$-function regularization scheme, the analytic extensions we have just computed are used to define the sum of the following divergent series:
\begin{align}
  1+1+1+\ldots&:=\zeta_R(0)=-\tfrac12\,,\\[2mm]
  1+2+3+\ldots&:=\zeta_R(-1)=-\tfrac1{12}\,,\\[2mm]
  1+4+9+\ldots&:=\zeta_R(-2)=0\,.
\end{align}
Certainly, any definition might seem uncanny insofar as it assigns a vanishing or even a negative value to a sum of positive terms. In section \ref{rama} --where we regularize these divergent series with a smooth function of compact support-- we will understand how these unexpected values appear after subtracting the infinite contribution originated by the removal of the cutoff. In appendix \ref{divsums} a comparison between analytic extensions and other definitions of divergent sums is sketched.

\section{Application: vacuum energy of a particle in a ring}\label{cas1}

Before giving an insight into this intriguing regularization of divergent sums, we apply the results they provide to the calculation of a physical quantity, namely the vacuum energy of a ring generated by the quantum oscillations of the field of a scalar particle. Later, in section \ref{cas2}, we will compute the vacuum energy in a more realistic setting where the predictions of the $\zeta$-function regularization have been experimentally confirmed.

Let us consider a massless scalar particle on a ring of radius $R$. Its dynamics in the context of relativistic quantum mechanics is described by a wave function $\phi(x,t)$ that satisfies
\begin{align}
  \left(\partial^2_t-\partial^2_x\right)\phi(x,t)=0\,.
\end{align}
This equation has infinitely many stationary solutions which can be identified with an integer $n\in\mathbb{Z}$,
\begin{align}\label{stat}
  \phi_n(x,t)\sim e^{ip_n x-i\omega_n t}\,,
\end{align}
where
\begin{align}
  \omega_n=+\sqrt{p_n^2}
\end{align}
and $p_n=n/R$, due to periodicity in $2\pi R$. From the point of view of classical mechanics, each of these solutions with momentum $n/R$ and positive energy $\omega_n$ is a travelling wave that can also be considered as a normal mode of frequency $\omega_n$ of a system of infinitely many coupled harmonic oscillators. The quantization of the field $\phi(x,t)$ is equivalent to the quantization of this infinite set of harmonic oscillators. As is well-known, due to quantum oscillations, the minimum energy of a single harmonic oscillator of frequency $\omega$ is given by $\frac12\hbar\omega$. As a consequence, the minimum energy of the quantized field should correspond to the (infinite) sum of the ground state energies of all the normal modes of the system of coupled oscillators.

In the language of quantum field theory, the space of physical states of the quantized field is spanned by vectors characterized by the number of scalar particles in each of the stationary solutions \eqref{stat}. In particular, there is a vector $|0\rangle$ which represents the vacuum state, i.e., the state that describes the system in the absence of such particles. Surprisingly, although the total number of particles in the vacuum state $\langle 0|\hat{N}|0\rangle$ vanishes, its energy $E_0=\langle 0|\hat{H}|0\rangle$ is not necessarily zero. As explained above, in the case of the particle in a ring, $E_0$ corresponds to the infinite sum
\begin{align}\label{ejuno}
  E_0&=\frac12\,\sum_{n\in\mathbb{Z}}\,\omega_n
    =\sum_{n=1}^\infty \sqrt{\frac{n^2}{R^2}}
    =\frac1R\,\sum_{n=1}^\infty n\,,
\end{align}
which --interestingly-- depends on the size of the ring. As a matter of fact, it is obvious from dimensional reasons that $E_0$ should be proportional to $1/R$ but we obtained for the proportionality coefficient a divergent series of positive terms; should it be a positive number? Is there a way to get some physical information from expression \eqref{ejuno}? In the $\zeta$-function regularization described in the previous section, the vacuum energy $E_0$ is defined as
\begin{align}
  E_0=\frac{1}{R}\,\zeta_R(-1)=-\frac1{12}\,\frac{1}{R}\,.
\end{align}
This result indicates that, even in the absence of the massless particle, the quantum oscillations of its vacuum state originate an attractive force under which the ring tends to shrink to zero size\footnote{This force could be considered as responsible for the smallness of compactified extra dimensions in Kaluza-Klein models \cite{Appelquist:1982zs}.}.

\section{Casimir force between two conducting plates}\label{cas2}

In this section we will determine the Casimir force between two parallel plates; our purpose is to give some more details of the basic framework of quantum field theory, as well as to illustrate some usual manipulations regarding the computation of the analytic extension of the $\zeta$-function in terms of the asymptotic expansion of the heat-trace.

Let us then consider a relativistic massive particle confined between two plates separated a distance $L$ in the $x$-direction. In the context of relativistic quantum mechanics the particle is described by a wave function $\phi(x,\vec{y},t)$ --with support in $0<x<L$, $\vec{y}\in\mathbb{R}^2$ and $t\in\mathbb{R}$-- that minimizes the action
\begin{align}\label{strip-action}
    S[\phi]=\int_{\mathbb{R}}dt\int_{\mathbb{R}^2}d^2y\int_0^L dx\ \tfrac12
    \left\{\dot{\phi}^2-|\nabla\phi|^2-m^2\phi^2\right\}\,.
\end{align}
These configurations therefore satisfy
\begin{align}\label{strip-eom}
  \left(\partial^2_t+A\right)\phi=0\,,
\end{align}
where $A$ is the positive, second order differential operator in $3$-dimensional space given by
\begin{align}
  A:=-\partial^2+m^2\,.
\end{align}
In addition, we impose some, say Dirichlet, boundary conditions at $x=0$ and $x=L$ that represent the effect of the plates. Let us compute the spectrum of this operator. The normalized eigenfunctions of $A$ are given by\footnote{These eigenfunctions satisfy the following orthogonality and completeness conditions:
\begin{align}
    \int_{\mathbb{R}^2} d^2y\int_0^L dx\ \phi^*_{\vec{k},n}(x,\vec{y})\phi_{\vec{k}',n'}(x,\vec{y})
    &=(2\pi)^2\,\delta(\vec{k}-\vec{k}')\,\delta_{nn'}\,,\nonumber\\
    \int_{\mathbb{R}^2} \frac{d^2k}{(2\pi)^2}\, \sum_{n=1}^\infty\
    \phi_{\vec{k},n}(x,\vec{y})\phi^*_{\vec{k},n}(x',\vec{y}')
    &=\delta(x-x')\,\delta(\vec{y}-\vec{y}')\,.\nonumber
\end{align}}
\begin{align}\label{nm}
  \phi_{\vec{k},n}(x,\vec{y})=\sqrt{\frac{2}{L}}\, \sin{(n\pi x/L)}\,e^{i\,\vec{k}\,\vec{y}}\,,
\end{align}
where $\vec{k}\in\mathbb{R}^2$ and $n\in\mathbb{Z}^+$; the corresponding eigenvalues can be written as $\omega^2_{\vec{k},n}$, with
\begin{align}\label{cf}
    \omega_{\vec{k},n}=\sqrt{\vec{k}^2+\frac{\pi^2 n^2}{L^2}+m^2}\,.
\end{align}
In consequence, stationary solutions to eq.\ \eqref{strip-eom} can be written as
\begin{align}\label{cm}
    \phi_{\vec{k},n}(x,\vec{y})\,e^{-i\omega_{\vec{k},n} t}\,.
\end{align}
Now, we are ready to obtain from the spectrum of the operator $A$ some physical information about the quantized field.

In quantum field theory, the scalar field is given by an operator built from a superposition of the normal modes \eqref{cm},
\begin{align}\label{thefield}
    \hat{\phi}(x,\vec{y},t)=\int_{\mathbb{R}^2} \frac{d^2k}{(2\pi)^2}\,\sum_{n=1}^\infty
    \ \frac{1}{2\omega_{\vec{k},n}}\left\{\hat{a}_{\vec{k},n}\,\phi_{\vec{k},n}\,e^{-i\omega_{\vec{k},n} t}
    +\hat{a}_{\vec{k},n}^\dagger\,\phi^*_{\vec{k},n}\,e^{i\omega_{\vec{k},n} t}\right\}\,.
\end{align}
Now, equal-time canonical commutation relations between $\hat{\phi}(x,\vec{y},t)$ and its conjugate field lead to the following algebra of creation and annihilation operators:
\begin{align}
    [\hat{a}_{\vec{k},n},\hat{a}_{\vec{k}',n'}^\dagger]
    =2\omega_{\vec{k},n}\ (2\pi)^2\, \delta(\vec{k}-\vec{k}')\,\delta_{nn'}\,.
\end{align}
This algebra can be represented in the Fock space generated by the repeated action of creation operators $\hat{a}_{\vec{k},n}^\dagger$ on the vacuum state $|0\rangle$, which is a vector annihilated by all $\hat{a}_{\vec{k},n}$.

Moreover, from the classical action \eqref{strip-action} one obtains the Hamiltonian of the quantized field, which corresponds --if written in terms of creation and annihilation operators-- to an infinite set of uncoupled harmonic oscillators (the normal modes described in section \ref{cas1}). In fact, the vacuum expectation value of this Hamiltonian (per unit area in the directions parameterized by $\vec{y}\in\mathbb{R}^2$) results
\begin{align}\label{naiveE}
    E_0=\langle 0|\hat H|0\rangle=\frac12\int_{\mathbb{R}^2} \frac{d^2k}{(2\pi)^2}
    \,\sum_{n=1}^\infty\ \omega_{\vec{k},n}\,,
\end{align}
which, as mentioned in section \ref{cas1}, is the sum of the ground state energies of an infinite set of harmonic oscillators with frequencies $\omega_{\vec{k},n}$.

Expression \eqref{naiveE} is of course ill-defined since the series and the integral are not convergent, due to the behaviour of $\omega_{\vec{k},n}$ for large $n$ and $|\vec{k}|$; this is a manifestation of the UV-divergencies one encounters when computing quantum corrections in field theories. The $\zeta$-function method provides a regularization of this divergence by means of the following function of the complex variable $s\in\mathbb{C}$:
\begin{align}\label{E}
    E_0(s)=\frac\mu2\int_{\mathbb{R}^2} \frac{d^2k}{(2\pi)^2}
    \,\sum_{n=1}^\infty \left(\frac{\omega_{\vec{k},n}}{\mu}\right)^{-2s}
    =\frac{1}{2}\,\mu^{2s+1}\,\zeta(s)\,,
\end{align}
where $\zeta(s)$ is the $\zeta$-function of the operator $A$, defined as
\begin{align}\label{z}
    \zeta(s)=\int_{\mathbb{R}^2}\frac{d^2k}{(2\pi)^2}\,\sum_{n=1}^\infty
    \ (\omega^2_{\vec{k},n})^{-s}
\end{align}
for $\mathcal{R}(s)>\tfrac32$, or as its analytic extension otherwise. Note that, for dimensional reasons, we have also introduced a parameter $\mu$ as an arbitrary scale with mass dimensions whose physical significance will be elucidated later. Finally, by comparison with eq.\ \eqref{naiveE}, the vacuum energy is defined as the analytic extension of $E_0(s)$ to $s=-\tfrac12$. If the $\zeta$-function were analytic at this point of the complex plane then the vacuum energy would be --as one would expect-- independent of the arbitrary scale $\mu$, but this might not be the case.

As we did in section \ref{ae} for the Riemann $\zeta$-function, in order to obtain the analytic extension of \eqref{z} we first compute the heat-trace of the operator $A$,
\begin{align}
    {\rm Tr}\,e^{-\tau A}
    &=\int_{\mathbb{R}^2}\frac{d^2k}{(2\pi)^2}\,\sum_{n=1}^\infty\ e^{-\tau\,\omega^2_{k,n}}
    =\int_{\mathbb{R}^2}\frac{d^2k}{(2\pi)^2}\,\sum_{n=1}^\infty
    \ e^{-\tau\left(\vec{k}^2+\frac{\pi^2 n^2}{L^2}+m^2\right)}\nonumber\\[2mm]
    &=\frac{e^{-\tau m^2}}{(4\pi\tau)^{\frac32}}
    \left\{L-\sqrt{\pi}\,\tau^{\frac12}+2L\, \sum_{n=1}^\infty e^{-\frac{L^2}{\tau}\,n^2}\right\}\,,
\end{align}
where, after integrating in $\vec{k}\in\mathbb{R}^2$, we have rewritten the sum by using Poisson inversion formula,
\begin{align}
  \sum_{n\in\mathbb{Z}} e^{-a n^2}=\sqrt{\frac{\pi}{a}}
  \ \sum_{n\in\mathbb{Z}} e^{-\frac{\pi^2}{a} n^2}
  \qquad ({\rm for\ }a>0)\,,
\end{align}
because it makes explicit the divergent behaviour of the heat-trace for small values of the parameter $\tau$. From relation \eqref{meltra} we obtain for the $\zeta$-function
\begin{align}\label{zetacomm}
    \zeta(s)
    &=\frac{1}{\Gamma(s)}\int_0^\infty d\tau\,\tau^{s-1}
    \ \frac{e^{-\tau m^2}}{(4\pi\tau)^{\frac32}}
    \left\{L-\sqrt{\pi}\,\tau^{\frac12}+2L\,\sum_{n=1}^\infty e^{-\frac{L^2}{\tau}\,n^2}\right\}
    \nonumber\\[2mm]
    &=\frac{m^{3-2s}L}{(4\pi)^{\frac32}}\,\frac{\Gamma(s-\tfrac32)}{\Gamma(s)}
    -\frac{m^{2-2s}}{8\pi}\,\frac{1}{s-1}\,
    +\frac{m^{\frac32-s}L^{s-\frac12}}{2\pi^{\frac32}\Gamma(s)}
    \,\sum_{n=1}^\infty \frac{K_{\frac32-s}(2mLn)}{n^{\frac32-s}}\,.
\end{align}
It is important to remark that the terms in the heat-trace that decrease exponentially as $\tau\rightarrow 0^+$ give contributions to the $\zeta$-function that are analytic in the whole complex plane. In conclusion, $\zeta(s)$ has a simple pole at $s=1$ (due to the second term in \eqref{zetacomm}) and an infinite number of other simple poles at $s=\frac32-\mathbb{Z}^+$ (due to the first term in \eqref{zetacomm}). In particular, it has a simple pole at $s=-\tfrac12$ with residue $-m^4L/32\pi^2$. The function $E_0(s)$ can thus be written at $s=-\tfrac12+\epsilon$ as
\begin{align}\label{anal-ext}
    E_0(-\tfrac12+\epsilon)=-\frac{m^4L}{64\pi^2}\ \frac{1}{\epsilon}+{\rm fin}(m,\mu,L)+O(\epsilon)\,,
\end{align}
where the finite part reads
\begin{align}
    {\rm fin}(m,\mu,L)&=\frac{m^3}{24\pi}+\frac{m^4L}{64\pi^2}\left[\log{\left(\frac{m^2}{\mu^2}\right)}
    +\frac12-2\log{2}\right]-\frac{m^2}{8\pi^2 L}\,\sum_{n=1}^\infty \frac{K_2(2mLn)}{n^2}\,.
\end{align}
As was already seen in expression \eqref{E}, a pole in $\zeta(s)$ at $s=-\frac12$ brings forth a (logarithmic) dependence of the vacuum energy on the arbitrary scale $\mu$. In the massless case, where the only dimensionful parameter is $L$, the analytic extension of $E_0(s)$ is finite at $s=-\tfrac12$ and thus $\mu$-independent; therefore, the limit $\epsilon\rightarrow 0$ of expression \eqref{anal-ext} gives an unambiguous result for the vacuum energy (per unit area),
\begin{align}\label{m=0}
  E_0=-\frac{\pi^2}{1440}\,\frac{1}{L^3}
  \qquad ({\rm for\ }m=0)\,.
\end{align}
However, the simple pole of $E_0(s)$ at $s=-\frac12$ for the massive case requires some additional physical prescription to define a renormalized vacuum energy. For instance, if the field is infinitely massive then quantum oscillations are blurred and, consequently, the vacuum energy is expected to vanish; thus an admissible procedure consists in removing (by a minimal subtraction) those terms in eq.\ \eqref{anal-ext} which do not vanish in the $m\rightarrow\infty$ limit. After this subtraction we obtain for the vacuum energy of the massive field
\begin{align}\label{m}
    E_0=-\frac{m^2}{8\pi^2 L}\,\sum_{n=1}^\infty \frac{K_2(2mLn)}{n^2}\,.
\end{align}
Taking into account that the physical consequences of the vacuum energy are associated with its variation with the separation of the plates, one could also remove from expression \eqref{anal-ext} those terms which are independent of $L$, as well as those linear in $L$ --which together with the vacuum energy in the space outside the plates give an $L$-independent contribution of the vacuum energy of the whole space $\mathbb{R}^3$. The result obtained with this prescription is again given by expression \eqref{m}.

\begin{figure}[t]
\centering
\begin{minipage}{.60\textwidth}
\includegraphics[height=50mm]{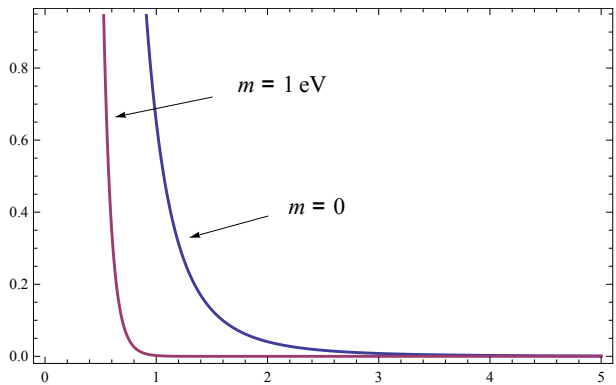}
\caption{Casimir attractive pressure (in mPa) as a function of the plates separation (in $\mu$m) for massless and massive ($m=1\,{\rm eV}$) fields.}
\label{casimir}
\end{minipage}
\end{figure}

In conclusion, there must exist an attractive force (per unit area) between two Dirichlet-type parallel plates, that is originated by the quantum oscillations of the vacuum state of the field. Since the force is exponentially decreasing for massive fields, the most prominent contribution corresponds to the oscillations of massless particles. The pressure on the parallel plates is given by the variation of the vacuum energy with their distance, which for the massless case reads (see  eq.\ \eqref{m=0}),
\begin{align}\label{pres}
  p:=-\partial_L E_0=-\hbar c\ \frac{\pi^2}{480}\,\frac{1}{L^4}\,.
\end{align}
This pressure is represented in figure \ref{casimir}, as well as for the massive case (see eq.\ \eqref{m}), as a function of the separation of the plates.
As the figure shows, at a plates separation of 1\,$\mu$m, massless fields originate a small pressure of the order of mPa; for a mass of $1\,{\rm eV}$ the pressure decreases to $0.3\%$ this value\footnote{Nevertheless, this exponential decrease of the Casimir force with the distance for massive fields is not verified, in general, in the presence of curved boundaries.}, whereas it is completely negligible for masses of the order of MeV.

This effect was predicted in 1948 by H.\ B.\ G.\ Casimir \cite{Casimir:1948dh}, who studied the attraction experienced by two parallel conducting plates due to the vacuum oscillations of the electromagnetic field\footnote{The vacuum pressure originally derived in \cite{Casimir:1948dh} is twice expression \eqref{pres} because of the two transverse polarizations of the electromagnetic field.}. Already in 1958 this Casimir force could be experimentally observed \cite{Sparnaay:1958wg}. Since then, much more accurate experiments have successfully measured the Casimir force in different settings \cite{Lamoreaux:1996wh,Mohideen:1998iz} and are used nowadays to test many theoretical models that take into account more realistic aspects, such as the finite conductivity of the boundaries, the roughness of their surfaces as well as the effect of temperature.

\section{Cutoff regularization}\label{rama}

In the previous sections we have defined the result of a divergent series in terms of the analytic extension of the $\zeta$-function and we have found that this result leads to physical predictions --the Casimir force-- that have been confirmed by very accurate experiments. Nevertheless, a physicist might be more familiar with the regularization of a divergent sum in terms of a cutoff function\footnote{Actually, H.\ B.\ G.\ Casimir used a cutoff function in \cite{Casimir:1948dh}, whose physical meaning he interpreted as the fact that for high energy virtual photons the conducting plates must become transparent.}. In this section we will compare both procedures in order to become more familiar with the meaning of the definition in terms of analytic extensions.

Based on Euler-Maclaurin formula
\begin{align}
  \sum_{n=1}^N f(n)&\sim \int_0^N dx\, f(x)+\frac12\left[f(N)-f(0)\right]
  +\sum_{n=1}^\infty \frac{B_{2n}}{(2n)!}\left[f^{(2n-1)}(N)-f^{(2n-1)}(0)\right]\,,
\end{align}
Ramanujan computed the value of a divergent series using an appropriate cutoff function \cite{Ramanujan}. We will depict his procedure by computing the same infinite sums which we have already computed using the (Riemann) $\zeta$-function regularization. Let then $\omega:\mathbb{R}^+\rightarrow \mathbb{R}$ be a smooth, positive and bounded function with support in the interval $[0,1]$ and such that, for some $\epsilon>0$, takes the constant value $\omega(x)=1$ for any $0\leq x\leq 1-\epsilon$. If we apply Euler-Maclaurin formula to the function $f(x)=x^k\,\omega(x/N)$, for $k\in\mathbb{Z}^+$, we readily obtain
\begin{equation}
  \sum_{n=1}^\infty n^k\,\omega(n/N)=N^{k+1}\,\int_0^\infty dx\, x^k\omega(x)\mp \frac{B_{k+1}}{k+1}\,,
\end{equation}
where the lower sign should be used only for the particular case $k=0$. We conclude that if we replace the infinite sum of the $k$-th powers of positive integers with the sum of the first $N$ terms --smoothed with a cutoff $\omega(n/N)$-- we obtain an expression which, unsurprisingly, diverges as $N\rightarrow\infty$. However, this divergence can be isolated in a term which corresponds to the integral in Euler-Maclaurin formula (proportional to the $k$-th moment $c_k[\omega]$ of $\omega(x)$). The remainder is a finite term which is independent of the cutoff and coincides with the value given by the analytic extension of the corresponding $\zeta$-function; Ramanujan claimed this term ``is like the center of gravity of a body''. In particular, for $k=0,1,2$, we obtain
\begin{align}
  \sum_{n=1}^\infty\, \omega(n/N)&\sim c_0[\omega]\,N-\frac12\,,\\
  \sum_{n=1}^\infty\, n\,\omega(n/N)&\sim c_1[\omega]\,N^2-\frac1{12}\,,\\
  \sum_{n=1}^\infty\, n^2\,\omega(n/N)&\sim c_2[\omega]\,N^3\,.
\end{align}
Note that after subtracting the leading (divergent) behaviour one obtains the values of the Riemann $\zeta$-function $\zeta_R(s)$ at $s=0,1,2$. From this point of view, the analytic extension $\zeta_R(s)$ gives the difference between the divergent ``smoothed'' series and the corresponding ``smoothed'' integral.

\section{Asymptotic expansion of the heat-trace}\label{ht}

Now we will state a very important mathematical result that describes the asymptotic expansion of the heat-trace ${\rm Tr}\,e^{-\tau A}$ for small values of $\tau$. We have already seen that this expansion allows the computation of the analytic extension of the $\zeta$-function; in particular, it provides complete information of its pole structure.

Let us consider a second order matricial differential operator $A$ acting on the space of $n$-component functions $\mathbb{C}^n\otimes L_2(\mathbb{R}^d)$. Such an operator defines a generalized heat-equation (or evolution equation in Euclidean time, if $A$ were a Hamiltonian)
\begin{align}
  \left(\partial_\tau+A\right)\,\phi(\tau;x)=0
\end{align}
whose solution --under some given initial condition $\phi(0;x)$-- can be written in terms of the integral operator $e^{-\tau\,A}$, called heat-operator (or evolution operator in Euclidean time), as follows,
\begin{align}
  \phi(\tau;x)=e^{-\tau\,A}\,\phi(0;x)=\int_{\mathbb{R}^d} dx'\,K(\tau;x,x')\,\phi(0;x')\,.
\end{align}
If, as before, $\mathbb{C}^n\otimes L_2(\mathbb{R}^d)$ has an orthonormal basis of eigenfunctions $\phi_1,\phi_2,\ldots$ of $A$ with eigenvalues $\lambda_1,\lambda_2,\ldots$, then the heat-kernel $K(\tau;x,x')$ can be written as the series
\begin{align}
  K(\tau;x,x')=\sum_{n=1}^\infty e^{-\tau\,\lambda_n}\ \phi_n(x)\,\phi_n^\dagger(x')\,,
\end{align}
and the heat-trace is
\begin{align}
  {\rm Tr}\,e^{-\tau\,A}=\int_{\mathbb{R}^d} dx\,{\rm tr}\,K(\tau;x,x)=\sum_{n=1}^\infty e^{-\tau\,\lambda_n}\,,
\end{align}
where ${\rm tr}$ represents the finite trace in the $\mathbb{C}^n$ part of $\mathbb{C}^n\otimes L_2(\mathbb{R}^d)$.

As the proper time $\tau$ tends to zero the heat-operator tends to the identity operator in an infinite dimensional Hilbert space, so the heat-trace diverges in this limit. The asymptotic behaviour of the heat-trace for small values of the proper time has important applications in QFT and has been systematically described for a quite general class of differential operators defined on smooth manifolds. It has been proved that, under quite general conditions, the following asymptotic expansion holds \cite{Vassilevich:2003xt}:
\begin{align}\label{he}
    {\rm Tr}\,e^{-\tau\,A}\sim \frac{1}{(4\pi\tau)^{d/2}}\,\sum_{n=0}^\infty a_n(A)\,\tau^{n}\,,
\end{align}
where the Seeley-de Witt coefficients $a_n(A)$ are integrals of local expressions of the coefficients of the differential operator $A$ (and the geometric properties of the base manifold in the case of curved spacetimes). If the base manifold has boundaries then the sum in \eqref{he} also includes positive half-integers $n\in\mathbb{Z}^+/2$. Nevertheless, in our applications we only need to consider the simpler case of the flat spacetime without boundaries $\mathbb{R}^d$.

Let $A$ be the second order differential operator
\begin{align}
    A=-(\mathbf{1}_n\,\partial_i+\omega_i(x))^2+V(x)\,,
\end{align}
where $\omega_i(x),V(x)$ are matrix-valued functions that take values in $\mathbb{C}^n\otimes \mathbb{C}^n$. Then the first Seeley-de Witt coefficients of expansion \eqref{he} are given by the following expressions:
\begin{align}
  a_0(A)&=\int dx\ {\rm tr}\,\mathbf{1}_n\,,\label{a0}\\
  a_1(A)&=-\int dx\ {\rm tr}\,V(x)\,,\label{a1}\\
  a_2(A)&=\int dx\ {\rm tr}\,\left\{\tfrac12\,V^2(x)-\tfrac16\,(\partial_i+\omega_i)^2\,V(x)
  +\tfrac1{12}\,(\partial_i\omega_j-\partial_j\omega_i+[\omega_i,\omega_j])^2\right\}\,.\label{a2}
\end{align}
Since we are considering a non-compact base manifold, these integrals are in general ill-defined; one should consider instead the expressions corresponding to \eqref{he} --as well as to \eqref{a0}, \eqref{a1} and \eqref{a2}-- for the local quantity $K(\tau;x,x)$ or, alternatively, for the trace of the heat-operator but including a convenient smearing function. Yet, as long as we appropriately take into account the infinite volume contributions, we will be able to derive from these expressions the effective action of a field theory over the whole $\mathbb{R}^d$.

The asymptotic expansion of the heat-trace, together with relation \eqref{meltra}, allows us to determine the pole structure of the $\zeta$-function by the same procedure followed in section \ref{ae} for the particular case of the Riemann $\zeta$-function,
\begin{align}\label{zetapolos}
  \zeta(s)&=\frac{1}{\Gamma(s)}\left\{\int_0^1 d\tau\,\tau^{s-1}\,
  \frac{1}{(4\pi\tau)^{d/2}}\,\sum_{n=0}^\infty a_n(A)\,\tau^{n}+{\rm (analytic\ terms)}\right\}\nonumber\\
  &=\frac{1}{\Gamma(s)}\left\{\sum_{n=0}^\infty \frac{a_n(A)}{(4\pi)^{d/2}}
  \ \frac{1}{s-(d/2-n)}+{\rm (analytic\ terms)}\right\}\,.
\end{align}
We conclude that, in general, the $\zeta$-function has simple poles at $d/2-\mathbb{Z}^+$ --except at negative integers, due to the cancellation with $1/\Gamma(s)$-- and that the residues are given by the Seeley-de Witt coefficients. Moreover, the value of $\zeta(s)$ at a negative integer $s=-k\in\mathbb{Z}^-$ is given by
\begin{align}
  \zeta(-k)=(-1)^k\,k!\,\frac{a_{d/2+k}(A)}{(4\pi)^{d/2}}\,,
\end{align}
for even $d$, but vanishes in odd dimensions.

\section{The effective action}\label{efac}

In a field theory described by the classical action $S[\phi]$, the effective action $\Gamma[\phi]$ is another functional of the field $\phi$ which --according to Feynman's quantization procedure-- is obtained after averaging all field configurations $\varphi$ with the weight $e^{-S[\varphi]}$. Since $S[\varphi]$ is measured in units of $\hbar$, the leading contribution to the effective action in an expansion in powers of $\hbar$ is given by the classical action. In this section we will show that the first quantum correction is given by the functional determinant of a differential operator determined by the second functional derivative of the classical action. We will then describe how the spectral functions are used to compute this functional determinant and we will apply the asymptotic expansion of the heat-trace to characterize the UV-divergent terms in the effective action. With this procedure we will finally compute the $\beta$-function of the self-coupling constant of a scalar field.

We begin by defining the partition function $Z[J]$ and the generating functional $W[J]$ --both depending on some arbitrary source $J(x)$-- as
\begin{align}\label{parfun}
  Z[J]:=e^{-\frac1\hbar\,W[J]}:=\int \mathcal{D}\varphi\ e^{-\frac1\hbar\,S[\varphi]+\frac1\hbar\int dx\,J\varphi}\,.
\end{align}
The mean field $\phi(x)$ (also understood as $\langle 0| \hat\varphi(x)|0\rangle$) in the presence of this source is
\begin{align}\label{fi}
  \phi(x):=\frac{1}{Z[J]}\ \int \mathcal{D}\varphi
  \ e^{-\frac1\hbar\,S[\varphi]+\frac1\hbar\int J\varphi}\, \varphi(x)
  =-\frac{\delta W[J]}{\delta J(x)}\,;
\end{align}
the functions $J(x)$ and $\phi(x)$ are thus called conjugate fields. Finally, we define the effective action as
\begin{align}
  \Gamma[\phi]:=W[J]+\int J\phi\,,
\end{align}
where $J$ must be understood as implicitly determined by $\phi$ through the relation \eqref{fi}. From this definition it immediately follows the equation
\begin{align}
  \frac{\delta \Gamma[\phi]}{\delta\phi(x)}=J(x)\,.
\end{align}
In consequence, in the absence of the ``external source'' $J(x)$, the vacuum expectation value $\phi(x)$ of the field minimizes the effective action $\Gamma$, unlike the classical configuration of the field, which minimizes the classical action $S$. Moreover, if we make a functional expansion of the effective action in powers of the field $\phi$, the coefficient of the $n$-th power of the field provides the (proper) scattering amplitude of $n$ particles. In this sense, the effective action describes the full behaviour of the system including all quantum effects.

Now it is convenient to consider $\phi(x)$ as an arbitrary configuration --determined by an appropriate $J(x)$-- and to shift the integration variable as $\varphi\rightarrow \phi+\varphi$ to obtain
\begin{align}
  Z[J]=e^{-\frac1\hbar\,S[\phi]+\frac1\hbar\int J\phi}
  \int \mathcal{D}\varphi
  \ e^{-\frac1\hbar
  \int (\delta_\phi S-J)\,\varphi-\frac1{2\hbar}\int\!\!\int\, \varphi\,\delta_\phi^2S\,\varphi+\ldots}\,,
\end{align}
where $\delta_\phi S$ denotes the functional derivative $\delta S[\varphi]/\delta \varphi(x)$ evaluated at the configuration $\phi(x)$. Similarly, the kernel $\delta_\phi^2S$ is given by $\delta^2 S[\varphi]/\delta\varphi(x)\delta\varphi(x')$ evaluated at $\phi(x)$. The effective action then reads
\begin{align}
  \Gamma[\phi]=S[\phi]-\hbar\log{\int \mathcal{D}\varphi
  \ e^{-\frac1{2}\int\!\!\int\, \varphi\,\delta_\phi^2S\,\varphi}
  \ e^{-\frac{1}{\sqrt{\hbar}}\int (\delta_\phi S-\delta_\phi \Gamma)\,\varphi+\sqrt{\hbar}\,O(\varphi^3)}}\,,
\end{align}
where we have rescaled $\varphi\rightarrow \sqrt{\hbar}\,\varphi$. Since $\Gamma=S+O(\hbar)$ we can write
\begin{align}
  \Gamma[\phi]=S[\phi]-\hbar\log{\int \mathcal{D}\varphi
  \ e^{-\frac1{2}\int\!\!\int\, \varphi\,\delta_\phi^2S\,\varphi}
  \,\left(1+O(\hbar)\right)}\,,
\end{align}
where $O(\hbar)$ comes from expressions which are quartic in $\varphi$. The Gaussian functional integral is given by the functional determinant of the operator $A$ which defines the quadratic form in the exponent, that is,
\begin{align}
  \Gamma[\phi]=S[\phi]+\frac{\hbar}2\log{{\rm Det}\,A}+O(\hbar^2)\,.
\end{align}
This expression gives the leading quantum contributions (one-loop corrections) to the effective action. The operator $A$, sometimes referred to as the operator of quantum fluctuations of the field, is defined by the kernel $\delta_\phi^2S$. Take, for example, the action of a real scalar field $\phi$ on (Euclidean) $\mathbb{R}^4$ with a quartic self-interaction
\begin{align}\label{quartic}
    S[\phi]=\int_{\mathbb{R}^4}dx\,\left\{
    \tfrac{1}{2}(\partial\phi)^2+\tfrac{1}{2}\,m^2\,\phi^2+\tfrac{1}{4!}\,\lambda\,\phi^4
    \right\}\,.
\end{align}
Then the operator of quantum fluctuations is the second order differential operator
\begin{align}\label{a-scalar}
  A=-\partial^2+m^2+\tfrac12\,\lambda\,\phi^2\,.
\end{align}
If we assume that $A$ has eigenvalues $\lambda_1,\lambda_2,\lambda_3,\ldots$, which in general grow to infinity, we must now indicate how to compute its determinant. The $\zeta$-function and the heat-trace provide different definitions of the functional determinant. In principle, we would expect
\begin{align}
  \log{{\rm Det}\,A}=\sum_n\log{\lambda_n}\,,
\end{align}
but this series is in general divergent. In the $\zeta$-function approach one introduces an appropriately decreasing power of the eigenvalues $\lambda_n^{-s}$ and then computes the analytic extension to $s=0$ of the regularized series,
\begin{align}
  \log{{\rm Det}\,A}:=\left.\sum_n\log{(\lambda_n/\mu^2)}
  \,\left(\lambda_n/\mu^2\right)^{-s}\ \right|_{s=0}\,.
\end{align}
The series is convergent for $\mathcal{R}(s)$ large enough and the determinant is defined as the analytic extension of this series to $s=0$. As before, we have introduced an arbitrary parameter $\mu$ with mass dimensions whose relevance in the determination of physical quantities will be discussed later. This definition can also be written as
\begin{align}
  \log{{\rm Det}\,A}=-\zeta'(0)-\zeta(0)\,\log{\mu^2}\,,
\end{align}
where
\begin{align}
  \zeta(s)=\sum_n\lambda_n^{-s}
\end{align}
is the $\zeta$-function of the operator $A$. As we have seen from \eqref{zetapolos}, the $\zeta$-function is analytic at $s=0$.

The heat-trace instead provides a different definition, based on the identity
\begin{align}
  \int_{\Lambda^{-2}}^\infty \frac{d\tau}{\tau}\,e^{-\tau \lambda_n}
  =\int_{\lambda_n/\Lambda^2}^\infty \frac{d\tau}{\tau}\,e^{-\tau}
  =-\log{(\lambda_n/\Lambda^2)}-\gamma+O(\lambda_ n/\Lambda^2)\,,
\end{align}
where $\gamma$ is Euler-Mascheroni constant and $\Lambda$ is some high-energy cutoff which could be eventually removed by making, when possible, $\Lambda\rightarrow\infty$. In this approach the determinant is then defined as
\begin{align}\label{det-ht}
  \log{{\rm Det}\,A}:=
  -\int_{\Lambda^{-2}}^\infty \frac{d\tau}{\tau}\,{\rm Tr}\,e^{-\tau A}\,.
\end{align}
In this case, the arbitrary mass scale is introduced by $\Lambda$. The limit $\Lambda\rightarrow\infty$ characterizes the UV-behaviour of the theory.

Let us now compute the one-loop effective action for the self-interacting scalar field in $\mathbb{R}^4$ described by the classical action \eqref{quartic}. If we use the heat-trace definition of the functional determinant we obtain
\begin{align}\label{effact-ht}
  \Gamma[\phi]=S[\phi]-\frac{1}2
  \int_{\Lambda^{-2}}^\infty \frac{d\tau}{\tau}
  \,e^{-\tau m^2}
  \,{\rm Tr}\,e^{-\tau \left\{-\partial^2+V(x)\right\}}\,,
\end{align}
where we have omitted $\hbar$. Note that we have factorized the mass term in $A$ (see \eqref{a-scalar}) from the heat-trace. The function $V(x)$ is the field-dependent expression
\begin{align}
    V(x)=\tfrac12\,\lambda\,\phi^2\,.
\end{align}
Next, we insert in \eqref{effact-ht} the heat-trace expansion as given by expression \eqref{he}, together with the expressions \eqref{a0}, \eqref{a1} and \eqref{a2} for the coefficients, to obtain
\begin{align}\label{effactphi4}
    &\Gamma[\phi]=S[\phi]-\frac{1}{2}\int_{\Lambda^{-2}}^\infty \frac{d\tau}{\tau}
    \ e^{-\tau m^2}\,\frac{1}{(4\pi\tau)^2}
    \left\{a_0+a_1\,\tau+a_2\,\tau^2+\ldots\right\}\nonumber\\
    &=S[\phi]-\frac{1}{32\pi^2}\int_{\Lambda^{-2}}^\infty d\tau\,\frac{e^{-\tau m^2}}{\tau^3}
    \left\{\int_{\mathbb{R}^4}dx\,1
    -\tau\int_{\mathbb{R}^4}dx\,\tfrac{\lambda}2\,\phi^2
    +\tau^2\int_{\mathbb{R}^4}dx\,\tfrac{\lambda^2}8\,\phi^4+\ldots\right\}\,.
\end{align}
In the last line we have only displayed the three terms that lead to divergencies in the effective action as $\Lambda\rightarrow \infty$: a field-independent term, a mass term (quadratic in the field) and a quartic self-interaction term. We therefore face two different problems: the dependence of the effective action on the cutoff $\Lambda$ and the infinities that appear when the cutoff is removed. Actually, these two unacceptable results are related and can be circumvented on the basis of a same assumption.

Note that --taking into account the cosmological constant-- these three types of terms are already present in the classical action. The renormalization procedure consists in regarding the original action $S$ in eq.\ \eqref{quartic}, as well as its parameters $m,\lambda$ (and the cosmological constant), as the result of integrating out (with some unknown, eventually more complex action) field configurations which, in Fourier space, have momenta above the arbitrary cutoff $\Lambda$. Thus, the action in expression \eqref{quartic} only describes field configurations with momenta below $\Lambda$. From this point of view, the parameters $m,\lambda$ clearly depend on $\Lambda$ and it is to be expected that this dependence is such that the effective action --from which physical results are obtained-- is finite and $\Lambda$-independent. Accordingly, the effective action can be written in terms of physical parameters as\footnote{In this model, after considering two loop corrections also the field $\phi$ must be renormalized.}
\begin{align}\label{physact}
    \Gamma[\phi]=\int_{\mathbb{R}^4}dx\,\left\{
    \tfrac{1}{2}(\partial\phi)^2+\tfrac{1}{2}\,m^2_{{\rm phys}}\,\phi^2
    +\tfrac{1}{4!}\,\lambda_{{\rm phys}}\,\phi^4\right\}\,,
\end{align}
where $m^2_{{\rm phys}}$ and $\lambda_{{\rm phys}}$ (of course, $\Lambda$-independent) are determined by physical measurements. Comparing expressions \eqref{effactphi4} and \eqref{physact} we readily obtain
\begin{align}
  m^2_{{\rm phys}}&=m^2\left\{1+\frac{\lambda}{32\pi^2}
  \int_{m^2/\Lambda^2}^\infty d\tau\ \frac{e^{-\tau}}{\tau^2}\right\}
  =m^2\left\{1+\frac{\lambda}{32\pi^2}
  \frac{\Lambda^2}{m^2}+\ldots\right\}\,,\\
  \lambda_{{\rm phys}}&=\lambda\left\{1-\frac{3\lambda}{32\pi^2}
  \int_{m^2/\Lambda^2}^\infty d\tau\ \frac{e^{-\tau}}{\tau}\right\}
  =\lambda\left\{1-\frac{3\lambda}{16\pi^2}\log{(\Lambda/m)}+\ldots\right\}\,.
\end{align}
These expressions intrinsically determine the dependence, up to one-loop corrections, of the original parameters $m,\lambda$ on the cutoff $\Lambda$. The $\beta$-function, given by the derivative of the coupling constant with respect to the cutoff $\Lambda$, then reads
\begin{align}
  \beta(\lambda):=\Lambda\,\partial_\Lambda\lambda=\frac{3}{16\pi^2}\,\lambda^2\,.
\end{align}
This expression is the leading quantum contribution; higher order quantum corrections contain higher powers of $\lambda$. Therefore, as long as the coupling constant is small, since its derivative is positive, it becomes larger as $\Lambda$ increases or, equivalently, as the theory involves field configurations with higher momenta. This justifies the use of perturbation theory to study the scattering amplitudes of these self-interacting scalar particles at low energies. However, if the coupling constant increases to an infinite value at some finite $\Lambda$ then the full theory becomes inconsistent, unless it is trivial; this is called Landau pole problem, and is also present in QED (see section \ref{sch}).

\section{Equivalence between both regularizations}\label{rela}

As we have seen, the $\zeta$-function and the heat-trace give two different definitions of the functional determinant. Let us briefly study how their difference arise in the computation of the effective action for a field theory described by an operator of quantum fluctuations $A$.

As shown in the previous section, it is convenient to separate the mass term from the operator $A$ so we assume in this section that the second functional derivative $\delta_\phi^2S$ is the kernel of the operator $A+m^2$. The functional determinant in the heat-trace approach reads
\begin{align}
  \log{{\rm Det}\,(A+m^2)}
  &=-\int_{\Lambda^{-2}}^\infty d\tau\ \frac{e^{-\tau m^2}}{\tau}\ {\rm Tr}\,e^{-\tau A}
  \nonumber\\
  &=-\frac{1}{(4\pi)^{d/2}}\,\sum_{n=0}^\infty m^{d-2n}\,\Gamma\left(-\tfrac{d}2+n,\tfrac{m^2}{\Lambda^2}\right)\ a_n(A)\,,
\end{align}
in terms of the incomplete gamma function. Let us analyse the UV-divergent terms in this series separately. For $0\leq n\leq d/2$ we get the sum\footnote{Strictly speaking, what follows holds for even $d$; for odd $d$ one must consider $0\leq n\leq d/2-1/2$.}
\begin{align}
  -\frac{1}{(4\pi)^{d/2}}\,\sum_{n=0}^{d/2} m^{d-2n}\,\Gamma\left(-\tfrac{d}2+n,\tfrac{m^2}{\Lambda^2}\right)\ a_n(A)\,,
\end{align}
which diverges as $\Lambda^{d-2n}$ for $\Lambda\rightarrow\infty$; the term corresponding to $n=d/2$ (only present in even dimensions) gives a logarithmically divergent contribution. These divergencies are removed by introducing similar counterterms in the classical action: if some of these terms are already present in the action then the corresponding constants get renormalized; otherwise, one says that new couplings are generated by quantum corrections. In any case, the theory can be made one-loop finite if the action contains or admits terms of the form given by $a_0(A),a_1(A),a_2(A),\ldots,a_{d/2}(A)$.

Once the divergencies in these terms are removed by a redefinition of the corresponding parameters in the classical action, the remaining terms in the functional determinant are given by
\begin{align}
  &-\frac{m^{-2}}{(4\pi)^{d/2}}\ \sum_{n=0}^\infty
  \ \frac{\Gamma\left(n+1\right)}{m^{2n}}\ a_{d/2+1+n}(A)
  &({\rm for\ even\ }d)\label{fincorr1}\\[2mm]
  &-\frac{m^{-1}}{(4\pi)^{d/2}}\ \sum_{n=0}^\infty
  \ \frac{\Gamma\left(n+1/2\right)}{m^{2n}}\ a_{d/2+1/2+n}(A)
  &({\rm for\ odd\ }d)\,.\label{fincorr2}
\end{align}

Let us now study the determinant as defined in the $\zeta$-function approach,
\begin{align}\label{detconm}
  \log{{\rm Det}\,(A+m^2)}=-\zeta'(0)-\zeta(0)\,\log{\mu^2}\,.
\end{align}
By means of the Mellin transform the $\zeta$-function can be written as
\begin{align}
  \zeta(s)
  &=\frac{1}{\Gamma(s)}\int_0^\infty d\tau\,\tau^{s-1}\,e^{-\tau m^2}
  \,\frac{1}{(4\pi\tau)^{d/2}}\,\sum_{n=0}^\infty a_n(A)\,\tau^n
  \nonumber\\
  &=\frac{m^{-2s}}{(4\pi)^{d/2}}\,\sum_{n=0}^\infty m^{d-2n}
  \,\frac{\Gamma(s-d/2+n)}{\Gamma(s)}\ a_n(A)\,.
\end{align}
Let us consider first the value of $\zeta(s)$ at $s=0$. For odd $d$ the term $1/\Gamma(s)$ cancels any contribution. In consequence, in odd dimensions $\zeta(0)=0$ and the determinant is independent of the arbitrary scale $\mu$. In even dimensions instead, the term $1/\Gamma(s)$ cancels the divergencies from the terms with $0\leq n\leq d/2$ at $s=0$ and the only non-vanishing contributions are
\begin{align}\label{zetacero}
  \zeta(0)
  =\frac{(-1)^{d/2}}{(4\pi)^{d/2}}\,\sum_{n=0}^{d/2}
  \,\frac{(-1)^n\,m^{d-2n}}{(d/2-n)!}\ a_n(A)\,.
\end{align}
Let us now compute the derivative of the $\zeta$-function,
\begin{align}\label{zetaprima}
  &-\zeta'(s)
  =\log{m^2}\,\zeta(s)-\mbox{}\nonumber\\
  &\mbox{}-\frac{m^{-2s}}{(4\pi)^{d/2}}\,\sum_{n=0}^\infty m^{d-2n}
  \,\frac{\Gamma(s-d/2+n)}{\Gamma(s)}
  \left\{\psi(s-d/2+n)-\psi(s)\right\}
  \, a_n(A) \,,
\end{align}
in terms of the gamma and digamma functions. At $s=0$ the first term is proportional to $\zeta(0)$ and together with the second term in the r.h.s.\ of \eqref{detconm} gives a total contribution $\log{(m^2/\mu^2)}\,\zeta(0)$ to the functional determinant. In consequence, the dependence on the arbitrary scale $\mu$ is proportional to the coefficients $a_0(A),a_1(A),\ldots,$ $a_{d/2}(A)$ and is removed from physical quantities by introducing an appropriate dependence on the scale $\mu$ of the original parameters in the action. This is equivalent to the dependence of the parameters on a renormalization scale $\mu$ when implementing renormalization prescriptions in terms of scattering amplitudes at some arbitrary scale of external momenta. On the other hand, the terms in $\zeta'(0)$ with $0\leq n\leq d/2-1$ give contributions to the effective action of the same form as the ones given in \eqref{zetacero}; thus they can also be considered as part of the counterterms that renormalize the classical action\footnote{Note that for $0\leq n\leq d/2-1$ the difference $\psi(s-d/2+n)-\psi(s)$ is analytic at $s=0$.}.

As regards the remaining terms in the series in \eqref{zetaprima}, in even dimensions the term $n=d/2$ vanishes, whereas the terms with $n> d/2$ give the series
\begin{align}
  -\frac{m^{-2}}{(4\pi)^{d/2}}\,\sum_{n=0}^\infty \frac{\Gamma(n+1)}{m^{2n}}\,a_{d/2+1+n}(A)\,,
\end{align}
which coincides with the heat-kernel result for the case of even $d$ (see \eqref{fincorr1}). Analogously, on can immediately prove that for odd $d$ we also get the same result as in the heat-kernel regularization.

In conclusion, both in the heat-trace and in the $\zeta$-function regularizations one introduces arbitrary scales, $\Lambda$ and $\mu$, respectively. A finite number of terms of the effective action diverge with the cutoff $\Lambda$ or depend on the arbitrary scale $\mu$; their dependence with the field is, in both cases, given by the first Seeley-de Witt coefficients $a_n$, with $0\leq n\leq d/2$. Therefore, the difference between both approaches lies in the dependence of the coupling constants with the large cutoff $\Lambda$ or the arbitrary scale $\mu$. This dependence is established through the experimental measurement of an appropriate number of scattering amplitudes; once this dependence is determined, the remaining contributions to the effective action (see eqs.\ \eqref{fincorr1} and \eqref{fincorr2}) are finite and unambiguously given by the Seeley-de Witt coefficients of higher order.

\section{Electrons in a constant magnetic field}\label{sch}

In this section we consider the coupling of electrons to a constant magnetic field and study, to leading order, the effect of the quantization of the electron field. Since fermions are quantized in terms of Grassmann variables, the one-loop contribution to the effective action of an electron/positron system in an electromagnetic background reads\footnote{Integration of Grassmann variables $\bar\psi,\psi$ satisfies $\int d\bar\psi\,d\psi\ e^{-a\bar\psi\psi}=a$.}
\begin{align}
  \Gamma=S-\log{\int \mathcal{D}\bar\psi\,\mathcal{D}\psi
  \ e^{-\int\, \bar\psi\,\left\{i\slashed{D}-m\right\}\,\psi}}
  =S-\log{{\rm Det} \left\{i\slashed{D}-m\right\}}\,,
\end{align}
where $S$ is the classical action for the electromagnetic field. We will use the heat-trace to compute this functional determinant. However, this method applies for positive operators so we need to perform another formal manipulation. Since in four dimensions there exists a charge conjugation matrix $C$ which satisfies $C\gamma^\mu C^{-1}=-(\gamma^\mu)^T$, we can write
\begin{align}
  \log{{\rm Det} \left\{i\slashed{D}-m\right\}}=\log{{\rm Det} \left\{-i\slashed{D}-m\right\}}\,.
\end{align}
As a consequence,
\begin{align}
  \Gamma&=S-\log{{\rm Det} \left\{i\slashed{D}-m\right\}}
  =S-\tfrac12\log{{\rm Det} \left\{\slashed{D}^2+m^2\right\}}\nonumber\\
  &=S+\frac12\int_{\Lambda^{-2}}^\infty \frac{d\tau}{\tau}\,e^{-\tau m^2}\ {\rm Tr}\, e^{-\tau\,\slashed{D}^2}\,.
\end{align}

The heat-trace in the case of an electron of charge $-e$ in the presence of a constant magnetic field $B>0$ (Landau problem) in Euclidean four-dimensional spacetime can be readily computed from the spectrum of the squared Dirac operator $\slashed{D}^2$ (see appendix \ref{landau-app}), which is given by
\begin{align}\label{lanlev}
  k^2+2eB\,(n+1/2)\pm eB\qquad{\rm with}\qquad k\in\mathbb{R}^2\qquad n=0,1,2,\ldots
\end{align}
The two-dimensional momentum $k$ corresponds to the timelike coordinate and the coordinate along the direction of the magnetic field. The second term in the eigenvalues corresponds to a one-dimensional harmonic oscillator of frequency $2eB$ in one of the transverse directions, whereas the last term is the coupling between the magnetic moment of the electron and the external field. Apart from the degeneracy $eB/2\pi$ per unit transversal area of these Landau levels there is an extra two-fold degeneracy corresponding to the electron/positron pair. From the spectrum \eqref{lanlev}, taking into account the degeneracies, the heat-trace results
\begin{align}
  {\rm Tr}\,e^{-\tau\,\slashed{D}^2}&=
  {\rm Vol(\mathbb{R}^4)}\ \frac{eB}{2\pi}\ \int_{\mathbb{R}^2}\frac{d^2k}{(2\pi)^2}\,e^{-\tau k^2}
  \ 2\,\sum_{n=0}^{\infty}\left\{e^{-\tau\,2eB\,(n+1)}+e^{-\tau\,2eB\,n}\right\}\nonumber\\
  &=\int_{\mathbb{R}^4}d^4x\ \frac{eB}{4\pi^2}\,\frac{\coth{(eB\tau)}}{\tau}
  \,.
\end{align}
We therefore obtain an explicit expression for the one-loop corrections to the effective action
\begin{align}\label{eab}
  \Gamma=S[B]+\frac{eB}{8\pi^2}\int_{\mathbb{R}^4}d^4x
  \int_{\Lambda^{-2}}^\infty \frac{d\tau}{\tau^2}\,e^{-\tau m^2}\,\coth{(eB\tau)}\,,
\end{align}
where the classical (Euclidean) action for the magnetic field reads
\begin{align}\label{eabc}
  S[B]=\int_{\mathbb{R}^4}d^4x\ \frac12\,B^2\,.
\end{align}
The divergence of the second term in expression \eqref{eab} as $\Lambda\rightarrow\infty$ can be removed upon the subtraction of the leading terms of the heat-trace for small $\tau$. These extra terms correspond to the renormalization of the cosmological constant and the electric charge. Let us write only the resulting finite effective action after the subtraction of these counterterms,
\begin{align}\label{effact-B}
  \Gamma[B]=\frac{eB}{8\pi^2}\int_{\mathbb{R}^4}d^4x
  \int_0^\infty \frac{d\tau}{\tau^2}\,e^{-\tau m^2}
  \left\{\coth{(eB\tau)}-\frac{1}{eB\tau}-\frac{eB\tau}3\right\}\,.
\end{align}
Since the expression between braces is $O(\tau^3)$ we could remove the cutoff $\Lambda$. The first counterterm gives a field-independent contribution to the effective action so, as mentioned, it is interpreted as a renormalization of the cosmological constant. On the other hand, the second term gives
\begin{align}
  \frac{eB}{8\pi^2}\int_{\mathbb{R}^4}d^4x \int_{\Lambda^{-2}}^\infty \frac{d\tau}{\tau}\,e^{-\tau m^2}
  \ \frac{eB\tau}3
  =\int_{\mathbb{R}^4}d^4x\ \frac12\left(\frac{e^2}{12\pi^2}
  \,\log{(\Lambda^2/m^2)}+\ldots\right)B^2\,,
\end{align}
where the dots denote terms which are finite as $\Lambda\rightarrow\infty$. This contribution represents a divergent correction to the classical action \eqref{eabc}, so it can be removed by a renormalization of the magnetic field,
\begin{align}
  B_{\rm phys}=\left(1+\frac{e^2}{24\pi^2}\,\log{\Lambda^2}+\ldots\right)B\,.
\end{align}
In order to mantain gauge invariance after quantum corrections the product of field and charge renormalizations must cancel to this order of perturbation theory; this gives the charge renormalization
\begin{align}
  e_{\rm phys}=e\left(1-\frac{e^2}{24\pi^2}\,\log{\Lambda^2}+\ldots\right)\,.
\end{align}
In consequence, the QED one-loop $\beta$-function reads
\begin{align}
  \beta(e)=\Lambda\,\partial_\Lambda e=\frac{e^3}{12\pi^2}\,.
\end{align}
As in the case of the $\lambda\,\phi^4$ theory considered in section \ref{efac}, the $\beta$-function is positive. This implies that the coupling constant increases with the cutoff and could eventually become infinite for a finite value of $\Lambda$ (Landau pole); equivalently, if the coupling constant $e$ is kept finite then the renormalized charge $e_{\rm phys}$ vanishes signaling a complete quantum screening of the electric charge (triviality problem). Although the $\beta$-function in this theory can only be computed perturbatively, there exists strong numerical evidence of the triviality of QED from nonperturbative lattice simulations. As a matter of fact, before 1970 the Landau pole problem seemed an unavoidable feature of any physically relevant field theory, until it was discovered that Yang-Mills theories have a negative $\beta$-function \cite{'tHooft:1971rn1,'tHooft:1971rn2,Politzer:1973fx,Gross:1973id}. We will carry out this calculation in section \ref{ym}.

Let us turn back to the finite expression \eqref{effact-B} for the one-loop corrections to the effective action of the magnetic field. In Minkowski space, the effective action is related to the vacuum persistence amplitude
\begin{align}
  e^{i\,\Gamma}=\langle 0;{\rm out}|0;{\rm in}\rangle\,,
\end{align}
so that
\begin{align}
    |\langle 0;{\rm out}|0;{\rm in}\rangle|^2=e^{-2\,\mathcal{I}(\Gamma)}\,.
\end{align}
This implies that if there is an imaginary part in the effective action in Minkowski spacetime then the vacuum is unstable and the probability of pair creation is not zero. However, the magnetic field is unchanged under an inverse Wick rotation so the effective action \eqref{effact-B} remains real in Minkowski spacetime and, as expected, there is no particle creation in a purely magnetic field background. Nevertheless, the situation changes in the presence of a constant electric field. The full expression in the presence of both electric and magnetic constant fields is called Heisenberg-Euler effective action \cite{Heisenberg:1935qt}. Since, in general, this action must depend on the invariants $E^2+B^2$ and $\vec{E}\cdot\vec{B}$, we can obtain the effective action for an electron/positron system in the presence of solely a constant electric field by simply replacing $B\rightarrow E$ in expression \eqref{effact-B}.

Now, upon an inverse Wick rotation, the electric field changes as $E\rightarrow -iE$ so (together with a global sign change) the effective action in a purely electric background in Minkowski spacetime reads
\begin{align}\label{elecmink}
  \Gamma[E]=-\frac{eE}{8\pi^2}\int_{\mathbb{R}^4}d^4x
  \int_0^\infty \frac{d\tau}{\tau^2}\,e^{-\tau m^2}
  \left\{\cot{(eE\tau)}-\frac{1}{eE\tau}+\frac{eE\tau}3\right\}\,.
\end{align}
Although no imaginary contribution is manifest in this expression, it is certainly ill-defined due to the singularities of the cotangent. Some prescription must be imposed in order to avoid these singularities; in other words, the integration path must be moved in the complex plane and thus imaginary contributions will appear leading to a nonvanishing probability of pair creation from the vacuum\footnote{This imaginary contribution can also be understood from the fact that the series expansion in powers of $eE$ is not Borel summable, as opposed to the purely magnetic case of expression \eqref{effact-B}, which leads to an alternating series in powers of $eB$ (see appendix \ref{divsums}).}. Since Euclidean time is defined in the negative imaginary axis, Euclidean electric field corresponds to an analytic extension to the positive imaginary axis. Therefore, to recover the effective action in Minkowski spacetime we must approach the positive real axis from above, which means that the electric field in \eqref{elecmink} has a small positive imaginary part\footnote{Equivalently, since the change $E\rightarrow -iE$ implies that eigenvalues that grow in the $+\infty$-direction turn into eigenvalues growing in the $-i\infty$-direction, one should introduce a positive imaginary part in the proper time $\tau$ in order to have a well-defined heat-trace in the definition of the functional determinant.},
\begin{align}
  \Gamma[E]&=-\frac{eE}{8\pi^2}\int_{\mathbb{R}^4}d^4x
  \int_{0}^{\infty} \frac{d\tau}{\tau^2}\,e^{-\tau m^2}
  \left\{\cot{[eE(1+i0)\tau]}
  -\frac{1}{eE\tau}+\frac{eE\tau}3\right\}\nonumber\\[2mm]
  &=-\frac{e^2E^2}{8\pi^2}\int_{\mathbb{R}^4}d^4x
  \int_{0}^{\infty} \frac{d\tau}{\tau^2}\,e^{-\tau \frac{m^2}{eE}}
  \left\{\cot{(\tau+i0)}
  -\frac{1}{\tau}+\frac{\tau}3\right\}\,.
\end{align}
This small imaginary part in the argument of the cotangent leads to a nonvanishing imaginary contribution due to the infinitely many simple poles at $\tau=\pi n$ (with $n=1,2,3,\ldots$) close to which
\begin{align}
  \mathcal{I}\left(\frac{1}{\sin{(\tau+i 0)}}\right)=
  \mathcal{I}\left(\frac{(-1)^n}{\tau-\pi n+i 0}\right)=
  -(-1)^n\pi\ \delta(\tau-\pi n)\,.
\end{align}
In consequence, the imaginary part of the effective action results
\begin{align}\label{prob}
  \mathcal{I}\left(\Gamma[E]\right)=\int_{\mathbb{R}^4}d^4x\ \left(\frac{e^2E^2}{8\pi^3}
  \,\sum_{n=1}^\infty \frac{e^{-\frac{\pi m^2}{eE}\,n}}{n^2}\right)\,.
\end{align}
This gives the pair production rate from vacuum induced by an external electric field, also known as Schwinger effect \cite{Schwinger:1951nm}. Expression \eqref{prob} shows that the probability of pair creation is exponentially suppressed for electric fields below the critical value
\begin{align}
  E_c\sim \frac{(mc^2)^2}{e}\,\frac{1}{\hbar\,c}\sim 1.3\cdot10^{18}\,{\rm volt/m}\,,
\end{align}
which is about three orders of magnitude above nowadays projected high-power laser facilities. Nevertheless, alternative theoretical settings which could enhance the rate of pair production are currently under study in order to estimate the possibility of reaching some experimental confirmation of this phenomenon in the next few years.

\section{$\beta$-function in pure Yang-Mills}\label{ym}

In this section we show how by direct application of the heat-trace expansion \eqref{he} one obtains the $\beta$-function in Yang-Mills theory. The non-abelian gauge field $A_\mu(x)=A^a_\mu(x)\,X^a$ takes values in the Lie algebra of some gauge group $G$ with basis vectors (in some convenient representation) $X^a$, $a=1,2,\ldots,{\rm dim}\,G$. The invariant classical action in four-dimensional pure Yang-Mills is given by
\begin{align}
  S[A]=\frac{1}{2g^2}\int_{\mathbb{R}^4} {\rm tr}\left(F_{\mu\nu}F_{\mu\nu}\right)
  =\frac{1}{4g^2}\int_{\mathbb{R}^4} F^a_{\mu\nu}F^a_{\mu\nu}\,,
\end{align}
where $g$ is the coupling constant,
\begin{align}
  F_{\mu\nu}=i[D_\mu,D_\nu]
\end{align}
is the field strength $F_{\mu\nu}(x)=F_{\mu\nu}^a(x)\,X^a$, and
\begin{align}
  D_\mu=\partial_\mu-i A_\mu\,.
\end{align}
In order to obtain the operator of quantum fluctuations, instead of computing the second functional derivative $\delta^2_A S$, we make the shift $A_\mu(x)\rightarrow A_\mu(x)+a_\mu(x)$ where now $A_\mu(x)$ is interpreted as a classical background field and the quantum gauge field is represented by $a_\mu(x)$ (Background Field Method). The second order expansion in $a_\mu(x)$ of the classical action then gives
\begin{align}\label{preac}
  S^{(2)}=\frac{1}{g^2}\int_{\mathbb{R}^4}dx\ {\rm tr}\left(
  a_\mu\left\{-[D_\nu,[D_\nu,a_\mu]]-[D_\mu,a_\mu][D_\nu,a_\nu]+2i\,a_\mu[F_{\mu\nu},a_\nu]\right\}\right)\,,
\end{align}
where the covariant derivative $D_\mu$ and the field strength $F_{\mu\nu}$ are now computed exclusively from the background $A_\mu$. The functional integral that gives the effective action contains a divergence --due to the integration along gauge equivalent field configurations-- which can be appropriately factorized by imposing a gauge condition on the quantum fields. If we choose the gauge condition $[D_\mu,a_\mu]=0$ and introduce in the quadratic part of the action the gauge fixing term in the Feynman gauge ($\xi=1$), then the second term in the r.h.s.\ of \eqref{preac} is cancelled and we get
\begin{align}
  S^{(2)}_{\rm gauge}=\frac{1}{g^2}\int_{\mathbb{R}^4}dx\ {\rm tr}\left(a_\mu\,\delta^2 S_{\rm gauge}\,a_\nu\right)\,,
\end{align}
where the operator of quantum fluctuations of the gauge field is now given by
\begin{align}
  \delta^2S_{\rm gauge}=-\delta_{\mu\nu}[D_\rho,[D_\rho,\cdot]]+2i\,[F_{\mu\nu},\cdot]
  =-\delta_{\mu\nu}\,(\partial-i A^{\rm adj})^2+2i\,F^{\rm adj}_{\mu\nu}\,.
\end{align}
On the other hand, after imposing the gauge choice $[D_\mu,a_\mu]=0$, gauge invariance is ensured by the introduction of (Grassmann) ghost fields with action
\begin{align}
  S_{\rm ghost}=\int_{\mathbb{R}^4}dx\ {\rm tr}\left(-\bar c(x)[D_\mu,[D_\mu,c(x)]]\right)\,.
\end{align}
The corresponding operator of quantum fluctuations of the ghost field then reads
\begin{align}
  \delta^2S_{\rm ghost}=-[D_\mu,[D_\mu,\cdot]]=-(\partial-i A^{\rm adj})^2\,.
\end{align}
Now, the one-loop effective action can be expressed in terms of the functional determinant of these operators,
\begin{align}\label{effact}
  \Gamma[A]=S[A]+\frac12 \log{\rm Det}\left\{\delta^2S_{\rm gauge}\right\}
  -\log{\rm Det}\left\{\delta^2S_{\rm ghost}\right\}\,.
\end{align}
There is no $-\frac12$ factor in the third term (as opposed to the second one) because, as already mentioned, the auxiliary ghosts are Grassmann fields (see also section \ref{sch}). The functional determinants in expression \eqref{effact} can be regularized by means of the heat-traces of the corresponding quantum fluctuation operators,
\begin{align}\label{effactlog}
  \Gamma[A]=S[A]-\frac12\int_{\Lambda^{-2}}^\infty \frac{d\tau}{\tau}\,e^{-m^2\tau}
  \left({\rm Tr}\,e^{-\tau\,\delta^2S_{\rm gauge}}-2\,{\rm Tr}\,e^{-\tau\,\delta^2S_{\rm ghost}}\right)\,.
\end{align}
Notice that we have also introduced an IR regulator $m$. If we replace in this expression the corresponding heat-trace expansions, the only divergent contributions arise from the terms proportional to the $a_2$ coefficients. According to \eqref{a2}, for the gauge field,
\begin{align}
  a^{\rm gauge}_2=\frac12\,4\,C_2(G)\,F^a_{\mu\nu}F^a_{\mu\nu}
  -\frac{1}{12}\,4\,C_2(G)\,F^a_{\mu\nu}F^a_{\mu\nu}
  =\frac53\,C_2(G)\,F^a_{\mu\nu}F^a_{\mu\nu}\,,
\end{align}
whereas for the ghost field,
\begin{align}
  a^{\rm ghost}_2=-\frac{1}{12}\,C_2(G)\,F^a_{\mu\nu}F^a_{\mu\nu}\,.
\end{align}
The Casimir factor $C_2(G)$ arises from the trace ${\rm tr}\,X^aX^b=C_2(G)\delta_{ab}$ in the adjoint representation. Replacing these contributions to the heat-traces into \eqref{effactlog} we obtain
\begin{align}\label{basta}
  \Gamma[A]&=S[A]-\frac12\int_{\Lambda^{-2}}^\infty \frac{d\tau}{\tau}\,e^{-m^2\tau}
  \,\frac{1}{(4\pi)^2}\int_{\mathbb{R}^4} dx\
  \left(a^{\rm gauge}_2-2\,a^{\rm ghost}_2\right)
  +{\rm (finite\ terms)}\nonumber\\
  &=S[A]-\frac{11}{96\pi^2}\,C_2(G)\,\log{(\Lambda/m)}\int_{\mathbb{R}^d} dx\ F^a_{\mu\nu}F^a_{\mu\nu}
  +{\rm (finite\ terms)}\,.
\end{align}
As is well-known, UV-divergencies in gauge theory are logarithmic. The divergence in \eqref{basta} can be removed by a redefinition of the coupling constant $g$ that appears in the classical action $S[A]$,
\begin{align}
  \frac{1}{g_{\rm phys}^2}=\frac{1}{g^2}\left(
  1-\frac{11}{24\pi^2}\,C_2(G)\,g^2\,\log{(\Lambda/m)}\right)\,.
\end{align}
The $\beta$-function then reads
\begin{align}
  \beta=\Lambda\partial_\Lambda g=-\frac{11}{24\pi^2}\,C_2(G)\,g^4\,.
\end{align}
Note that, being $\beta<0$, if the coupling constant is small at large energies then it decreases as $\Lambda\rightarrow\infty$ so that there is no Landau pole and perturbative calculations become appropriate for describing short-distance processes. This behaviour of the coupling constant is consistent with the anti-screening effect of virtual gluons.

\section{Quantum Field Theory at finite temperature}\label{fintem}

The partition function $Z[J]$ given by expression \eqref{parfun} can also be used to determine the thermodynamical properties of quantum fields at finite temperature. The standard procedure consists in performing an analytic continuation to Euclidean time $t$ in the operator of quantum fluctuations and imposing periodic boundary conditions at $t=0$ and $t=\beta$, where $\beta$ stands for the inverse temperature. In order to illustrate this method we consider the simplest example, namely, a non-interacting scalar particle of mass $m$ confined in a one-dimensional box of length $L$. After implementing the Wick rotation, the partition function (for $J(x)=0$) reads
\begin{align}\label{partemp}
  \log{Z}&=-\frac12\,\log{\rm Det}\left\{-\partial^2_t-\partial^2_x+m^2\right\}\nonumber\\
  &=\frac12\,\zeta'(0)+\frac12\,\zeta(0)\,\log{\mu^2}\,,
\end{align}
where the $\zeta$-function is given by
\begin{align}\label{hadwe}
  \zeta(s)
  &=\sum_{n=1}^\infty\sum_{k\in\mathbb{Z}}
  \left\{\left(\frac{2\pi k}{\beta}\right)^2+\left(\frac{\pi n}{L}\right)^2+m^2\right\}^{-s}\nonumber\\
  &=\frac{1}{\Gamma(s)}\int_0^\infty d\tau\,\tau^{s-1}\,e^{-\tau m^2}\,\sum_{n=1}^\infty\sum_{k\in\mathbb{Z}}
  e^{-\frac{4\pi^2 k^2}{\beta^2}\,\tau}\,e^{-\frac{\pi^2 n^2}{L^2}\,\tau}\,.
\end{align}
Note that the first term in the eigenvalues corresponds to the oscillating modes along the compactified Euclidean timelike direction (Matsubara frequencies). If we consider fermionic particles, we should impose antiperiodic boundary conditions at $t=0$ and $t=\beta$, instead. The second term in the eigenvalues comes from imposing Dirichlet boundary conditions at $x=0$ and $x=L$. In the second line of \eqref{hadwe} we have applied, as usual, the Mellin transform that relates the $\zeta$-function to the heat-trace.

\begin{figure}[t]
\centering
\includegraphics[height=60mm]{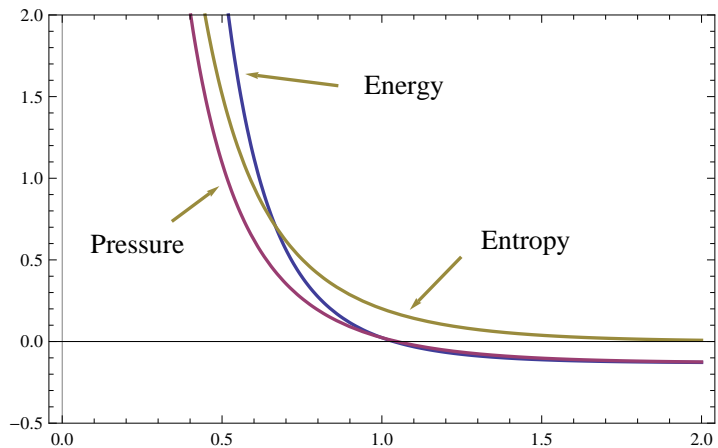}
\caption{Mean energy, pressure and entropy for a scalar massless particle in a one-dimensional box of unit length.}
\label{thermo}
\end{figure}

Next, we use Poisson inversion formula in the sum over $k\in\mathbb{Z}$ and afterwards, for the resulting terms corresponding to $k=0$, we use Poisson inversion in the sum over $n=1,2,3,\ldots$ After all these manipulations, the result reads
\begin{align}
  \zeta(s)
  &=-\frac{\beta}{4\sqrt{\pi}\Gamma(s)}\int_0^\infty d\tau\,\tau^{s-\frac32}\,e^{-\tau m^2}
  +\frac{\beta L}{4\pi\Gamma(s)}\int_0^\infty d\tau\,\tau^{s-2}\,e^{-\tau m^2}
  +\mbox{}\nonumber\\
  &\mbox{}+\frac{\beta L}{2\pi\Gamma(s)}\int_0^\infty d\tau\,\tau^{s-2}\,e^{-\tau m^2}
  \,\sum_{n=1}^\infty e^{-\frac{L^2}{\tau}\,n^2}
  +\mbox{}\nonumber\\
  &\mbox{}+\frac{\beta}{\sqrt{\pi}\Gamma(s)}\int_0^\infty d\tau\,\tau^{s-\frac32}\,e^{-\tau m^2}
  \,\sum_{n=1}^\infty\sum_{k=1}^\infty
  e^{-\frac{\beta^2}{4\tau}\,k^2}\,e^{-\frac{\pi^2 n^2}{L^2}\,\tau}\,.
\end{align}
Finally, we perform all integrations in $\tau$ to get
\begin{align}
  &\zeta(s)
  =-\frac{\beta m^{1-2s}\,\Gamma(s-\tfrac12)}{4\sqrt{\pi}\,\Gamma(s)}
  +\frac{\beta L m^{2-2s}}{4\pi\,(s-1)}
  +\frac{\beta L^s m^{1-s}}{\pi\Gamma(s)}\,\sum_{n=1}^\infty \frac{K_{1-s}(2Lmn)}{n^{1-s}}
  +\mbox{}\nonumber\\
  &\mbox{}+2^{\frac32-s}\frac{\beta^{s+\frac12}}{\sqrt{\pi}\Gamma(s)}
  \,\sum_{n=1}^\infty\sum_{k=1}^\infty\left(m^2+\frac{\pi^2 n^2}{L^2}\right)^{\frac14-\frac{s}2}\,k^{s-\frac12}
  \,K_{\frac12-s}\left(\beta k\sqrt{m^2+\frac{\pi^2 n^2}{L^2}}\right)\,.
\end{align}
Due to the $1/\Gamma(s)$ factor, the function $\zeta(s)$ at $s=0$ is given exclusively by the second term
\begin{align}
  \zeta(0)=-\frac{\beta L m^{2}}{4\pi}\,.
\end{align}
On the other hand, the derivative $\zeta'(s)$ at $s=0$ reads\footnote{$K_{1/2}(z)=\sqrt{\frac{\pi}{2z}}\,e^{-z}$.}
\begin{align}
  \zeta'(0)&=\frac{m\beta}2+\frac{\beta L m^{2}}{4\pi}\,\left(-1+\log{m^2}\right)
  +\frac{\beta m}{\pi}\,\sum_{n=1}^\infty \frac{K_{1}(2Lmn)}{n}
  +\mbox{}\nonumber\\
  &\mbox{}+2\,\sum_{n=1}^\infty\sum_{k=1}^\infty\frac{e^{-\beta k\sqrt{m^2+\frac{\pi^2 n^2}{L^2}}}}{k}\,.
\end{align}
We can now use expression \eqref{partemp} to evaluate the partition function\footnote{$\sum_{k=1}^\infty \frac{x^k}{k}=-\log{(1-x)}$.},
\begin{align}\label{part}
  \log{Z}&=\frac{m\beta}4+\frac{\beta L m^{2}}{8\pi}\,\left(-1+\log{(m^2/\mu^2)}\right)
  +\frac{\beta m}{2\pi}\,\sum_{n=1}^\infty \frac{K_{1}(2Lmn)}{n}
  +\mbox{}\nonumber\\
  &\mbox{}-\sum_{n=1}^\infty \log{\left(1-e^{-\beta \sqrt{m^2+\frac{\pi^2 n^2}{L^2}}}\right)}\,.
\end{align}
This expression depends on the arbitrary scale $\mu$ --due to the nonvanishing value of $\zeta(0)$-- and must be thus appropriately redefined by means of some renormalization prescription, as the vanishing of $\log{Z}$ for infinite values of $m^2$. This determines the partition function for a massive scalar field confined in the interval $[0,L]$. For simplicity, we consider next the massless case
\begin{align}
  \log{Z}=\frac{\pi}{24 L}\,\beta
  -\sum_{n=1}^\infty \log{\left(1-e^{-\beta \frac{\pi n}{L}}\right)}\,.
\end{align}
Note that the first term is proportional to the vacuum energy
\begin{align}
  E_0=\frac12\,\left.\sum_{n=1}^\infty \left(\frac{\pi n}{L}\right)^{-s}\ \right|_{s=-1}
  =\frac{\pi}{2L}\,\zeta_R(-1)=-\frac{\pi}{24L}\,,
\end{align}
whereas the second term is the partition function for the grand canonical ensemble. We can now compute, at temperature $\beta$, the mean energy
\begin{align}
  E:=-\partial_\beta \log{Z}
  =-\frac{\pi}{24 L}+\frac{\pi}{L}\,\sum_{n=1}^\infty\ \frac{n}{e^{\beta \frac{\pi n}{L}}-1}\,,
\end{align}
the pressure
\begin{align}
  p:=\frac{1}{\beta}\,\partial_L \log{Z}
  =-\frac{\pi}{24 L^2}+\frac{\pi}{L^2}\,\sum_{n=1}^\infty\ \frac{n}{e^{\beta \frac{\pi n}{L}}-1}\,,
\end{align}
and the entropy
\begin{align}
  S:=\log{Z}+\beta\,E
  =-\sum_{n=1}^\infty \log{\left(1-e^{-\beta \frac{\pi n}{L}}\right)}
  +\frac{\pi}{L}\,\sum_{n=1}^\infty\ \frac{n}{e^{\beta \frac{\pi n}{L}}-1}\,.
\end{align}
These thermodynamic quantities are represented in figure \ref{thermo} as a function of the inverse temperature $\beta$.
At high temperatures (small $\beta$) all these three quantities diverge, whereas at zero temperature ($\beta\rightarrow \infty$) the entropy vanishes, as expected, but the pressure and the mean energy attain a finite non-zero value corresponding to the vacuum oscillations of the quantum field.

\section{Noncommutative quantum fields}\label{nc}

Our last application of functional methods in quantum field theory is related to the computation of $n$-point functions from the Seeley-de Witt coefficients. In particular, we will compute one-loop corrections to the 2-point function in a noncommutative field theory in order to show the so-called UV/IR mixing effect.

Noncommutative field theories are formulated on a spacetime whose coordinates satisfy
\begin{align}
  [\hat x_i,\hat x_j]=2i\,\Theta_{ij}\,,
\end{align}
where $\Theta$ is an $\mathbb{R}^{4\times 4}$ antisymmetric matrix with dimensions of squared length. The fields are thus operators $\hat\phi(\hat x)$ that depend on these noncommutative coordinates. By means of the Weyl-Wigner transformation this space of noncommutative fields is isomorphic to the algebra of ordinary functions $\phi(x)$ (defined on the usual commutative $\mathbb{R}^4$) with the following noncommutative $\star$-product:
\begin{align}
  \phi(x)\star\chi(x):=
  \phi(x)\,e^{i\,\Theta_{ij}\,\overleftarrow{\partial}_i\overrightarrow{\partial}_j}\,\chi(x)\,.
\end{align}
Under this nonlocal associative product, know as Moyal-product, ordinary coordinates satisfy $x_i\star x_j-x_j\star x_i=2i\,\Theta_{ij}$. We will consider the case where $\Theta$ has maximal rank and that, after an appropriate choice of coordinates, can be written as
\begin{align}
  \Theta=\left(\begin{array}{cccc}
  0&\theta&0&0\\-\theta&0&0&0\\0&0&0&\theta\\0&0&-\theta&0\\
  \end{array}\right)\,;
\end{align}
$\theta\in\mathbb{R}^+$ is now the parameter that characterizes noncommutativity. In this formulation the Euclidean action of a noncommutative scalar field with a quartic self-interaction reads
\begin{align}
  S[\phi]=\int_{\mathbb{R}^4} dx\,\left\{
  \tfrac12\,(\partial\phi)^2+\tfrac12\,m^2\,\phi^2+\tfrac1{4!}\,\lambda\,\phi_\star^4
  \right\}\,,
\end{align}
where $\phi^4_\star:=\phi\star\phi\star\phi\star\phi$. Since the $\star$-product and the ordinary commutative product differ in total derivatives, quadratic terms in the action do not need to be modified in Moyal space.

After computing the second functional derivative, the field fluctuation operator results
\begin{align}\label{nc-a}
  \delta^2S=-\partial^2+m^2+\frac\lambda{3!}\,
  \left\{L(\phi^2)+R(\phi^2)+L(\phi)R(\phi)\right\}\,,
\end{align}
where $L$ and $R$ represent (left- and right-) Moyal-multiplication. In consequence, the one-loop effective action is given by
\begin{align}\label{nceffact}
    \Gamma[\phi]=S[\phi]-\frac{1}{2}\int_0^\infty \frac{d\tau}{\tau}
    \ \,e^{-\tau m^2}\,{\rm Tr}\,e^{-\tau\,\left\{
    -\partial^2
    +\tfrac{\lambda}{3!}\,L(\phi^2)+
    \tfrac{\lambda}{3!}\,R(\phi^2)+
    \tfrac{\lambda}{3!}\,L(\phi)R(\phi)
    \right\}}\,.
\end{align}
Since the field fluctuation operator is nonlocal, the standard heat-trace asymptotics described in section \ref{ht} no longer holds \cite{Fursaev:2011zz}. Let us study the effects of nonlocality on the quantum propagator (or 2-point function), which is given by the part of the effective action that is quadratic in the field $\phi$.

As can be seen from \eqref{nceffact}, there are three contributions to the effective action which are quadratic in the field arising from each term $L(\phi^2)$, $R(\phi^2)$ or $L(\phi)R(\phi)$ in an $O(\lambda)$ expansion of the exponential. The contribution of the first two of these terms is the same as in the commutative case: it is a divergent quantity that is regularized upon a mass renormalization. The most remarkable consequence of nonlocality on the heat-trace is due to the last term in the operator $\delta^2S$, which mixes left- and right- Moyal multiplication. As we will see, the contribution of the term $L(\phi)R(\phi)$, though UV-finite, leads to an IR-divergence in higher-order quantum corrections. In the Feynman diagram formulation of noncommutative field theories this term corresponds to a one-vertex nonplanar diagram. In order to evaluate this contribution we need the following result:
\begin{align}
  {\rm Tr}\,\left\{e^{-\tau\,\left(-\partial^2\right)}
  L\left(\phi\right)R\left(\phi\right)\right\}
  &=\frac{1}{(4\pi\tau)^2}\int_{\mathbb{R}^4} dx\,
  \phi(x)\,e^{-\frac{1}{\tau}\,(\Theta^2)_{ij}\,\partial_i\partial_j}\,\phi(x)\nonumber\\
  &=\frac{1}{(4\pi\tau)^2}\int_{\mathbb{R}^4} \frac{d^4p}{(2\pi)^4}\,
    \tilde\phi^*(p)\tilde\phi(p)\,e^{-\frac{\theta^2}{\tau}\,p^2}
  \,,
\end{align}
which has been obtained in \cite{Bonezzi:2012vr} using the worldline formalism. In the second line we have used $\Theta^2=-\theta^2\,\mathbf{1}_4$. In consequence, the nonplanar contribution to the quadratic effective action reads
\begin{align}
   \Gamma^{(2)}_{{\rm NP}}[\phi]&=\frac{1}{2}\int_{\mathbb{R}^4} \frac{d^4p}{(2\pi)^4}\,
    \tilde\phi^*(p)\tilde\phi(p)\ \frac{\lambda}{3!}
    \int_0^\infty d\tau\,\frac{e^{-\tau m^2}}{(4\pi\tau)^2}\,e^{-\frac{\theta^2}{\tau}\,p^2}\nonumber\\
    &=\frac{1}{2}\int_{\mathbb{R}^4} \frac{d^4p}{(2\pi)^4}\,
    |\tilde\phi(p)|^2\ \frac{\lambda m}{48\pi^2}\,\frac{K_1(2m\theta|p|)}{\theta|p|}\,.
\end{align}
This expression shows that the one-loop propagator contains a contribution
\begin{align}
   \Gamma^{(2)}_{{\rm NP}}(p,p')&=
   \frac{\pi^2}{3}\,\lambda m\,\delta(p+p')\,\frac{K_1(2m\theta|p|)}{\theta|p|}\,,
\end{align}
which is finite but diverges for small values of the momentum, even for massive fields. As a consequence, although this contribution is UV-finite, the use of this propagator in higher-order calculations (where $p$ represents the momentum of virtual particles and is thus off-shell) introduces IR-divergencies. This phenomenon, known as UV/IR mixing \cite{Minwalla:1999px}, can be interpreted in the following way. Note that the $\star$-product can be written as
\begin{align}
  (\phi\star\phi)(x)=\frac{1}{{\rm det}\,\Theta}
  \int_{\mathbb{R}^8}dx'\,dx''\ \phi(x')\phi(x'')
  \ e^{-i\,\Theta^{-1}_{ij}\,(x'-x)_i(x''-x)_j}\,.
\end{align}
Therefore, if the support of $\phi(x)$ is restricted to a distance $\delta_1$ in the $x_1$-direction, then $\phi^2_\star$ is nonzero in a distance $\delta_2\sim\theta/\delta_1$ in the $x_2$-direction. This means that a wave packet of width $\delta_1\ll\sqrt{\theta}$ gives a function $\phi^2_\star$ of width $\delta_2\gg \sqrt\theta$. In other words, small pulses are largely spread upon self-interaction, even for massive fields --which have exponentially decreasing propagators. As a consequence, scattering at low energies is affected by high energy virtual particles.

\section*{Acknowledgements}
I heartily thank hospitality at Centro de Estudios en F\'\i sica y Matem\'aticas B\'asicas y Aplicadas at UNACH. I also acknowledge the students and colleagues at CEFyMAP and IFLP who attended this series of talks which are part of the project ME/13/16 financially supported by CONACYT (M\'ex.) and MINCyT (Arg.)

\appendix

\section{Divergent sums}\label{divsums}

In this appendix we will illustrate with examples some criteria which are used to define divergent sums. In order to make contact with analytic extensions it is preferable to work with alternating sums, so we define
\begin{align}
  \eta_R(s)=1^{-s}-2^{-s}+3^{-s}-4^{-s}+\ldots
\end{align}
if $\mathcal{R}(s)>1$, or as its analytic extension otherwise. Its relation with the Riemann $\zeta$-function can be derived from the expression
\begin{align}
  1^{-s}-2^{-s}+3^{-s}-\ldots&=(1^{-s}+2^{-s}+3^{-s}+\ldots)-\mbox{}\nonumber\\
  &\mbox{}-2\times 2^{-s}(1^{-s}+2^{-s}+3^{-s}+\ldots)\,,
\end{align}
which, extended to the whole complex plane, gives
\begin{align}
  \eta_R(s)=(1-2^{1-s})\,\zeta_R(s)\,.
\end{align}
From the values of $\zeta_R(s)$ at $s=0,-1,-2$ --computed in section \ref{ae}-- we get the following values of the $\eta$-function:
\begin{align}
  \eta_R(0)&=\tfrac12=1-1+1-\ldots\\[2mm]
  \eta_R(-1)&=\tfrac14=1-2+3-\ldots\\[2mm]
  \eta_R(-2)&=0=1-4+9-\ldots
\end{align}
We will next compare this definition of the alternating divergent sums in terms of the analytic extension of the $\eta$-function with some well-known criteria.

Let us first consider Ces\`aro summation which, if the sequence of partial sums does not converge, defines the divergent series as the limit, if it exists, of the averages of the divergent sums. Take, for example, the series $1-1+1-\ldots$ whose partial sums give $1,0,1,0,\ldots$ Although this oscillating series is not convergent the averages of the partial sums give the sequence $1,\frac12,\frac23,\frac12,\frac35,\ldots$ that converges to $\frac12$. Therefore, we write
\begin{align}\label{c1}
  1-1+1-\ldots=\frac12\qquad (C,1)\,,
\end{align}
where $(C,1)$ stands for ``the first average of partial sums''. To explain this notation, let us next consider the series $1-2+3-\ldots$ whose partial sums give $1,-1,2,-2,\ldots$. The sequence of averages of this sequence is $1,0,\frac23,0,\frac35,0,\ldots$ which oscillates around the two values $0$ and $\frac12$. However, a new average of these averages gives the sequence $1,\frac12,\frac59,\frac5{12},\ldots$ which converges to $\frac14$. Therefore, we write
\begin{align}\label{c2}
  1-2+3-\ldots=\frac14\qquad (C,2)\,,
\end{align}
where $(C,2)$ stands for the second average which was required to consider in order to obtain a convergent sequence. Following the same arguments one can prove that
\begin{align}\label{c3}
  1-4+9-\ldots=0\qquad (C,3)\,.
\end{align}
As expected, the more divergent the absolute values of the term in the series are, the more number of averages have to be taken into account. The interpretation of this method is that if a series $\sum a_n$ is not summable, then the sequence $a_1,a_1+a_2,a_1+a_2+a_3,\ldots$ is not convergent. The average of partial sums instead give the sequence $a_1,a_1+\frac12\,a_2,a_1+\frac23\,a_2+\frac13\,a_3,\ldots$ The idea behind $(C,1)$ is that instead of adding a full term $a_n$ at each step of a partial sum, only a fraction of it is added; of course, this fraction becomes closer to one as the number of steps increases. As one can easily verify, a higher number of averages means that the fraction of a given term which is added at each partial sum approaches $1$ slower.

It can be proved that $(C,n+1)$ is stronger than $(C,n)$, meaning that if a series can be summed according to $(C,n)$ then it can also be summed (to the same value) according to $(C,n+1)$. In general, Ces\`aro summation is linear in the series, regular (it is stronger than the usual criterion for convergent series) and stable (if one removes the first term one gets a summable series which differs from the original one in this term).

Let us next consider Abel criterion, which defines the value of a series $\sum a_n$ as
\begin{align}
  \sum_{n=1}^\infty a_n
  :=\lim_{x\rightarrow 1^-}\ \sum_{n=1}^\infty a_n\,x^n\,,
\end{align}
as long as the r.h.s.\ converges for $x<1$ and the limit exists. Note that this method coincides with a heat-kernel type regularization --based on the cutoff function $e^{-t\,n}$-- if we define $x:=e^{-t}$. It can be shown that Abel criterion is linear, stable and stronger than Ces\`aro criterion. For example, taking the limit $x\rightarrow 1^-$ in the convergent expansions
\begin{align}
  \frac{1}{1+x}&=1-x+x^2-\ldots\\[2mm]
  \frac{1}{(1+x)^2}&=1-2x+3x^2-\ldots\\[2mm]
  \frac{1-x}{(1+x)^3}&=1-4x+9x^2-\ldots
\end{align}
one obtains the same values as in \eqref{c1}, \eqref{c2} and \eqref{c3}.

Finally, let us consider Borel summation, which is stronger than Abel criterion. To understand Borel summation it is convenient to write
\begin{align}
  \sum_{n=1}^\infty a_n=\sum_{n=1}^\infty\frac{a_n}{n!}\,n!
  =\sum_{n=1}^\infty\frac{a_n}{n!}\int_0^\infty d\tau\,\tau^n\,e^{-\tau}
\end{align}
and then formally interchange the sum with the integral to define
\begin{align}
  \sum_{n=1}^\infty a_n:=\int_0^\infty d\tau\,e^{-\tau}\ \sum_{n=1}^\infty\frac{a_n}{n!}\,\tau^n\,.
\end{align}
One can verify that Borel summation also gives the values in \eqref{c1}, \eqref{c2} and \eqref{c3}. The strength of this method can be understood in the divergent series $\sum(-1)^n\,n!\,g^n$, for positive $g$. Borel summation readily gives
\begin{align}\label{borel}
  \sum_{n=1}^\infty (-1)^n\,n!\,g^n=\frac{1}{g}
  \int_0^\infty d\tau\ \frac{e^{-\tau/g}}{1+\tau}\,,
\end{align}
which leads to the result $0!-1!+2!-\ldots=0.596347\ldots$ The importance of expression \eqref{borel} is that it indicates that a divergent series which is Borel summable might occur in a physical problem as the result of expanding a non-perturbative expression.

In general, a series $\sum a_n\,g^n$ is Borel summable only if $\sum a_n\,\tau^n/n!$ converges in some neighbourhood of the origin to some function $\mathcal B(\tau)$ that can be analytically extended to the positive real axis. Then $\sum a_n\,g^n$ is defined as the integral $g^{-1}\int_0^\infty e^{-\tau/g} \mathcal B(\tau)\,d\tau$, as long as this integral exists. In fact, under certain assumptions on analyticity and on the decrease of the remainder, Watson's theorem states that an asymptotic series uniquely determines the sum in terms of an integral of Borel's type \cite{Watson}.

It it interesting to apply this definition to expressions \eqref{effact-B} and \eqref{elecmink} for the effective action of an electron/positron system in constant magnetic and electric backgrounds, respectively. In the purely magnetic case, the function $\mathcal B(\tau)$ is given by $(\coth{\tau}-1/\tau-\tau/3)/\tau^2$, which is analytic in a neighbourhood of the positive real axis and --upon integration in $\tau$-- leads to an alternating series (in powers of $g=eB/m^2$) for the effective action which is Borel summable. On the contrary, for the purely electric case, $\mathcal B(\tau)$ is given by the function $(\cot{\tau}-1/\tau+\tau/3)/\tau^2$, which has infinitely many simple poles in the positive real axis and, if integrated in $\tau$, leads instead to a divergent series (in powers of $g=eE/m^2$) of positive terms. In the presence of singularities some prescription must be implemented, as was shown in section \ref{sch}, in order to perform the $\tau$-integral \cite{Dunne:2004nc}.

\section{Landau problem}\label{landau-app}

Let us consider a massless Dirac spinor of charge $-e$ in the presence of a constant magnetic field $B$ in the positive $z$-direction. The corresponding Dirac operator is
\begin{align}\label{dirope}
  \slashed{D}=\gamma^{\mu}(\partial_\mu-ie A_\mu)\,,
\end{align}
with $A^\mu=(0,0,Bx,0)$. In our notation $x^\mu=(t,x,y,z)$. We use the metric $\eta^{\mu\nu}=(+---)$ and, as a representation of the Clifford algebra $\{\gamma^\mu,\gamma^\nu\}=2\eta^{\mu\nu}$, the Weyl basis:
\begin{align}
  \gamma^0=\left(\begin{array}{cc}0&\mathbf{1}_2\\\mathbf{1}_2&0\end{array}\right)\,,
  \qquad
  \gamma^i=\left(\begin{array}{cc}0&\sigma^i\\-\sigma^i&0\end{array}\right)\,.
\end{align}
The squared operator
\begin{align}\label{D^2}
  \slashed{D}^2&=\left(\partial-ieA\right)^2-\tfrac14[\gamma^\mu,\gamma^\nu]\,ie\,F_{\mu\nu}\nonumber\\
  &=\Box-2ieBx\partial_y+e^2B^2x^2+eB\left(\begin{array}{cc}\sigma^3&0\\0&\sigma^3\end{array}\right)
\end{align}
restricted to the subspace
\begin{align}
  \Psi(t,x,y,z)=\left(\begin{array}{c}\phi(x)\\ \chi(x)\end{array}\right)
  e^{-ik_0t-ik_2y-ik_3z}
\end{align}
can be written as
\begin{align}
  \slashed{D}^2=-k_0^2+k_3^2-\partial_x^2+e^2B^2(x-k_2/eB)^2
  +eB\left(\begin{array}{cc}\sigma^3&0\\0&\sigma^3\end{array}\right)  \,.
\end{align}
Therefore, in Euclidean spacetime ($k_0\rightarrow ik_0$) the spectrum is
\begin{align}
  k_0^2+k_3^2+2eB(n+1/2)\pm eB
\end{align}
with $k_0,k_3\in\mathbb{R}$, $n=0,1,2,\ldots$ and an extra double degeneracy corresponding to the two Weyl spinors. There is also an infinite degeneracy because the eigenvalues do not depend on $k_2$. However, we can measure this degeneracy in terms of the area $L_xL_y$ transversal to the magnetic field. The shift $k_2/eB$ in the $x$-direction imposes the constraint $|k_2|/eB\leq L_x/2$. Since the number of states corresponding to this range of momenta is $2\times eB\times L_x/2\times L_y/2\pi$ we obtain a total degeneracy $L_xL_y\times eB/2\pi$.

Let us next consider the massless Dirac spinor of charge $-e$ in the presence of a constant electric field in the $x$-direction. Using $A^\mu=(-Ex,0,0,0)$ for the Dirac operator defined in \eqref{dirope} we obtain
\begin{align}
  \slashed{D}^2=\Box+2ieEx\partial_0-e^2E^2x^2+ieE\left(\begin{array}{cc}\sigma^1&0\\0&-\sigma^1\end{array}\right)
  \,,
\end{align}
which restricted to the subspace
\begin{align}
  \Psi(t,x,y,z)=\left(\begin{array}{c}\phi(x)\\ \chi(x)\end{array}\right)
  e^{-ik_0t-ik_2y-ik_3z}
\end{align}
reads
\begin{align}
  \slashed{D}^2=k_2^2+k_3^2-\partial_x^2-e^2E^2(x-k_0/eE)^2
  +ieE\left(\begin{array}{cc}\sigma^1&0\\0&-\sigma^1\end{array}\right)\,.
\end{align}
Note that the operator contains an inverted harmonic oscillator. Tunneling across this barrier gives the pair production rate. Alternatively, the introduction of a positive imaginary part in the proper time provides the exponential decrease that allows the use of heat-kernel techniques.

%%%%%%%%%%%%%%%%%%%%%%%%%%%%%%%%%%%%%%%%%%%%%%%%%%%%%%%%%%%%

\end{document}